\documentclass[preprint,5p,authoryear,a4paper,10pt]{elsarticle}
\pdfoutput=1
\usepackage[colorlinks=true,citecolor=blue,urlcolor=blue,linkcolor=blue]{hyperref}
\usepackage{amsfonts,amssymb,amsmath}
\usepackage{isomath}
\usepackage{siunitx}
\newcommand{\los}{line-of-sight}
\newcommand{\fov}{field-of-view}
\newcommand{\corotating}{co-rotating}
\newcommand{\spatiotemp}{spatio-tem\-poral}
\newcommand{\ie}{{\it i.e.}{\ }}

\DeclareSIUnit\Rsun{\ensuremath{R_\odot}}
\DeclareSIUnit\cmc{\per\centi\meter\tothe{3}}

\DeclareMathOperator{\argmin}{argmin} % use like: \underset{x}{\argmin}
\providecommand{\argminover}[1]{\underset{#1}{\argmin}\,}
\newcommand{\bvec}{\vectorsym}
\newcommand{\mtx}{\matrixsym}
\providecommand{\transp}[1]{#1^\mathsf{T}}
\DeclareMathOperator{\diag}{diag}
\providecommand{\norm}[1]{\left\lVert#1\right\rVert}

\newenvironment{tolerant}[1]{%
  \par\tolerance=#1\relax
}{%
  \par
}

\begin{document}

\begin{frontmatter}

\title{Time-Dependent Tomographic Reconstruction  of the Solar Corona}
\author[lam]{D.Vibert\corref{cor1}}
\ead{didier.vibert@lam.fr}

\author[lam]{C.Peillon}
\ead{christelle.peillon@lam.fr}

\author[lam]{P.~Lamy}
\ead{philippe.lamy@lam.fr}

\author[umich]{R.A.~Frazin}
\ead{rfrazin@umich.edu}

\author[lam]{J.~Wojak}
\ead{julien.wojak@lam.fr}

\cortext[cor1]{Corresponding author}
\address[lam]{Aix Marseille Universit\'e, CNRS, LAM (Laboratoire d'Astrophysique de Marseille) UMR 7326, 13388, Marseille, France}
%\address[lam]{Laboratoire d'Astrophysique de Marseille, UMR 7236, CNRS \& Aix-Marseille Universit\'e, 38 rue Fr\'ed\'eric Joliot-Curie,\ 
%13388 Marseille cedex 13, France}
\address[umich]{Dept. of Atmospheric, Oceanic and Space Sciences, University of Michigan, 
  Ann Arbor, MI 48109, USA}

%=====================================================================================================================================
\begin{abstract}
%=====================================================================================================================================

Solar rotational tomography (SRT) applied to white-light coronal images observed at multiple aspect angles has been the preferred approach for determining the three-dimensional (3D) electron density structure of the solar corona.
However, it is seriously hampered by the restrictive assumption that the corona is time-invariant which introduces significant errors in the reconstruction.
We first explore several methods to mitigate the temporal variation of the corona by decoupling the ``fast-varying'' inner corona from the ``slow-moving'' outer corona using multiple masking (either by juxtaposition or recursive combination) and radial weighting.
Weighting with a radial exponential profile provides some improvement over a classical reconstruction but only beyond $\approx$ \SI{3}{\Rsun}.
We next consider a full time-dependent tomographic reconstruction involving \spatiotemp{} regularization and further introduce a co-rotating regularization aimed at preventing concentration of reconstructed density in the plane of the sky.
%We also consider the introduction of an additional prior, a minimum background corona.
Crucial to testing our procedure and properly tuning the regularization parameters is the introduction of a time-dependent MHD model of the corona based on observed magnetograms to build a time-series of synthetic images of the corona.
Our procedure, which successfully reproduces the time-varying model corona, is finally applied to a set of of 53 LASCO-C2 $pB$ images roughly evenly spaced in time from 15 to 29 March 2009.
Our procedure paves the way to a time-dependent tomographic reconstruction of the coronal electron density to the whole set of LASCO-C2 images presently spanning 20 years.

\end{abstract}

%\keywords{Solar corona, Electron density, Tomography}

\end{frontmatter}

%==================================================================================
\section{Introduction}
%==================================================================================

Progress in understanding the physics of the corona and the solar wind depends upon empirical constraints on the three-dimensional (3D) distributions of the plasma properties such as temperature, density and magnetic field in the Sun's corona. 
Solar rotational tomography (SRT) uses coronal images observed at multiple aspect angles thanks to solar rotation to provide 3D reconstructions of the corona assuming that it does not vary in time, at least during the time interval necessary to achieve a complete view (typically half a solar rotation). 
When the images are at soft X-ray or EUV wavelengths, one can reconstruct both the temperature and the density in 3D from about \SIrange{1.03}{1.25}{\Rsun} (the lower bound is greater than \SI{1.0}{\Rsun} due to optical depth effects encountered in some spectral lines), via a process called differential emission measure tomography (DEMT), see \citet{Frazin09} and \citet{Barbey13}. 
When the input images are coronagraphic white-light images of the K-corona, one may reconstruct the density at greater heights. 
For example, when the image data are from the LASCO-C2 coronagraph, the reconstruction region is roughly from \SIrange{2.4}{5.5}{\Rsun} \citep{Frazin02} and when the source is STEREO/COR1 the reconstruction region is between about \SIrange{1.5}{4.0}{\Rsun} \citep{Kramar09,Kramar14}. 
In addition, the use of white-light coronagraph data has the potential of taking advantage of both the polarized brightness ($pB$) and total brightness ($B$) as independent sources of information \citep{Frazin10}.

Whereas the determination of the 3D distribution of the coronal density (as well as temperature) is a highly desirable objective, SRT has fundamental limitations that have prevented it from becoming broadly accepted as a useful tool for solar science. 
Calibration uncertainty in the image data is important \citep{Frazin12}, the finite \fov{} of the imaging instrument causes ``pile-up'' artifacts near the upper boundary \citep{Frazin02,Frazin10}, and the $\approx \SI{7}{\degree}$ tilt of the Sun's rotational pole relative to the ecliptic is theoretically a source of non-uniqueness, but the dominant source of error and uncertainty is the inherently dynamic nature of the solar corona so that it evolves as the Sun rotates. 
Thus, when one sees a movie of the corona, it is difficult to disentangle effects of the coronal dynamics from those of solar rotation, leading to a fundamental ambiguity and very serious artifacts in tomographic reconstructions of the corona as first shown by \citet{Frazin05a}. 

Since then, a new method to perform time-dependent tomography involving Kalman filters was developed by \mbox{\citet{Butala10}} and applied to STEREO/COR1 images. 
It did improve the quality of the reconstruction by reducing the number of artifacts; however, and as we shall discuss later on, it faced several difficulties leaving the time-dependent tomography problem un\-der-de\-ter\-mi\-ned and the solution very reliant on the regularization choices.
Moreover the complexity of the Kalman filter turns out to be of marginal utility due to the fact that we only have uninformative dynamical models of the corona at our disposal, whereas a time-dependent tomography can be achieved by a simpler multiple regularization approach as proposed in this article.
Interestingly, the EUV tomography problem appears to be less compromised by this issue, most likely due to the fact that lines-of-sight that hit the solar disk (as well as those above the limb, as in the case of coronagraph data) are included in the inversion, which tend to stabilize the solution. 
Thus, to date, most scientific results from SRT have been produced with EUV tomography \citep{Vasquez09,Vasquez10,Vasquez11,Huang12,Nuevo13}.

This article introduces several new strategies intended to mitigate the undesirable effects of coronal dynamics on tomographic reconstructions based on white-light coronagraph images.  
These new methods involve two different approaches:
\begin{enumerate}
\item For the part of the reconstruction located at a given radius $r_0$ and above, the goal is to de-emphasize the influence of coronal dynamics taking place at inferior radii $(r < r_0)$.  
The rationale behind this is that, at large heights, the corona is less dense and tends to be less dynamic.  
Thus, it is hoped that the reconstruction at $r > r_0$ will not be ``contaminated'' by the contribution of material with \los{} at $r < r_0$. 
These procedures involve weighting and masking schemes.
Part of the justification for this approach is that, strictly speaking, one does not need projections inferior to $r_0$ to reconstruct the part of the object at $r_0$ and above \citep{Frazin10,Louis83}.

\item In using Kalman filters or \spatiotemp{} regularization for tomography of the solar corona, it has been observed that, unless very strong temporal regularization is used (greatly reducing the temporal variability of the solution), the reconstructed density tends to be concentrated in the plane that contains the Sun's center and is perpendicular to the \los{} of the observation that was made at time $t$. 
Thus, the solution tracks the observation angle as the Sun rotates. 
The new strategy introduced to mitigate this effect is a novel type of \spatiotemp{} regularization, not based on gradients (as are most regularization operators), but instead specifically designed to suppress this rotational mode in the solution.
\end{enumerate}
The lessons learned from these new strategies are likely applicable to tomography based on EUV images, but that is not explored here.

The article is organized as follows: first, a general explanation of the tomographic reconstruction of the solar corona is given in Section~\ref{theory}. 
Then improved static reconstructions are described in Section~\ref{static}. 
Next, time-dependent tomographic reconstruction methods are described in Section~\ref{dynamic} and applied to model and real images. 
Finally, we conclude in Section~\ref{conclusion}.

%==================================================================================
\begin{tolerant}{6000}
\section{Tomographic Reconstruction of the Solar Corona: General Considerations}
\end{tolerant}
\label{theory}
%==================================================================================

\subsection{Formulation of SRT as a Linear Problem}
\label{formulation}
\begin{figure} % 1
\begin{center}
 \includegraphics[width=\columnwidth]{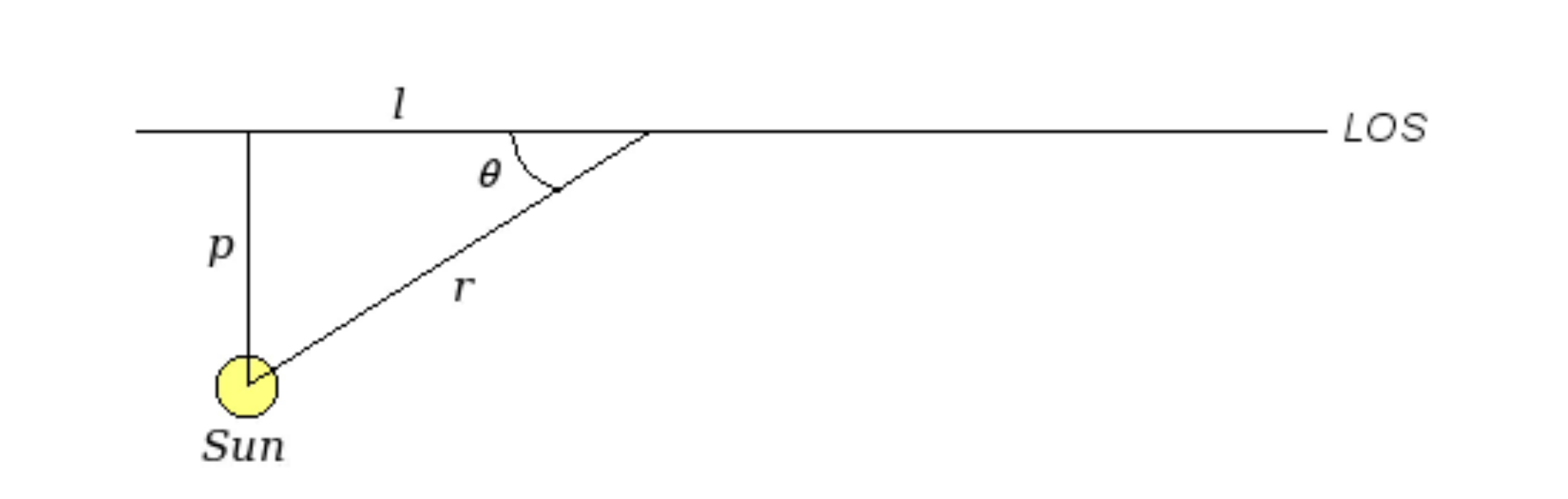}
\caption{The geometry for the light scattering in the solar corona}
\label{los}
\end{center}
\end{figure}

%-------------------------------------------------- 
We present below the formulation of solar rotational tomography appropriate to the case of white-light images obtained with coronagraphs.
Figure~\ref{los} illustrates the geometry for the Thompson scattering by a small volume of electrons located at a distance $r$\ from the Sun and the integration along the \los{} (denoted LOS thereafter) defined by its distance of closest approach ($p$) or impact parameter that produces the coronal radiance in a given pixel of the image.
The small volume under consideration is at distance $l$ from the point of closest approach of the LOS and it is convenient to define a scattering angle $\sin \theta = p/r$.

Following \citet{VdH50}, we can relate the local electron density ($N_{e}$) to the radiance measured in the $j^{th}$ pixel of a coronal image via the integral along the LOS
\begin{equation}
 y_{j} = \int^{\infty}_{-\infty} w[r(l_{j}),\theta (l_{j})]N_{e}[r(l_{j})]\,{\mathrm d}l_{j} \,,
 \label{VdH}
\end{equation}
where the Thomson scattering function ($w$) depends upon the type of images, either unpolarized radiance ``$B$'' or polarized radiance ``$pB$'', whose intensities are denoted as $w_{B}$ and $w_{pB}$, respectively.
Although the work here uses the more general expressions that include the finite size of the solar disk and limb darkening \citep{VdH50}, for simplicity we reproduce the approximate expressions given by \citet{Frazin10}, which are valid for a point-source of luminosity $4 \pi L$: 

\begin{equation}
 w_{B} = \frac{3\sigma_{e}}{8}\frac{L}{p^{2}} [ \sin^{2}\theta(l_{j})-\frac{1}{2}\sin^{4}\theta(l_{j}) ] \,,
 \label{wB}
\end{equation}
\begin{equation}
 w_{pB} = \frac{3\sigma_{e}}{16}\frac{L}{p^{2}} \sin^{4}\theta(l_{j}) \,,
 \label{wpB}
\end{equation}
where $\sigma_{e}$ is the Thomson cross section. 

In order to reconstruct the corona, we need to observe it during at least one half of a solar rotation, that is 14 days, with a minimum cadence of about one image per day, but a higher cadence is highly desirable.
Time intervals larger than one-half of a rotation are problematic for static reconstruction methods, creating many artifacts, because coronal dynamics causes inconsistency in the data.   
Then, for each pixel of each image, we have an equation along the corresponding LOS of the form of Equation~\eqref{VdH}. 
The resulting system of equations is solved by adopting a discrete representation of $N_{e}$ on a spherical grid \citep{Frazin09}. 
Replacing the integrals by summations over discrete intervals, we now have to solve a system of linear equations:
\begin{equation}
\bvec{y}=\mtx{A}\bvec{x}+\bvec{n} \,,
\label{syseq}
\end{equation}
where $\bvec{x}$ is a vector describing the discrete representation of $N_{e}$, $\bvec{y}$ is a vector containing the radiance values in all the pixels of the images, $\mtx{A}$ is a matrix calculated with Equation~\eqref{wB} or Equation~\eqref{wpB}, and $\bvec{n}$ is a Gaussian additive noise. 

Let $\mtx{N}$ be the covariance matrix of this noise on the observed pixels. 
The maximum likelihood solution to Equation~\eqref{syseq} is the least-squares solution:
\begin{equation}
\bvec{\hat{x}}= \argminover{\bvec{x}}
\transp{(\bvec{y}-\mtx{A}\bvec{x})}\mtx{N}^{-1}(\bvec{y}-\mtx{A}\bvec{x}) \,.
\label{mlew}
\end{equation}

Assuming an independent and identically distributed (IID) Gaussian noise, the matrix $\mtx{N}$ is diagonal, isotropic and assumes the form $\mtx{N} = \sigma^{2} \mtx{I}$. Then Equation~\eqref{mlew} becomes:
\begin{equation}
\bvec{\hat{x}}= \argminover{\bvec{x}}\norm{
\bvec{y}-\mtx{A}\bvec{x}}_2^2 \,.
\label{mle}
\end{equation}

If $\mtx{A}$ is not rank-deficient (more unknowns than independent equations), the analytic solution to Equation~\eqref{mlew} is given by:
\begin{equation}
 \bvec{\hat{x}}=[\transp{\mtx{A}}\mtx{N}^{-1}\mtx{A}]^{-1}\transp{\mtx{A}}\mtx{N}^{-1}\bvec{y}
\label{mlewsol}
\end{equation}
which, in the case of IID noise, simplifies to the solution of Equation~\eqref{mle}:
\begin{equation}
 \bvec{\hat{x}}=[\transp{\mtx{A}}\mtx{A}]^{-1}\transp{\mtx{A}}\bvec{y} \,.
\label{mlesol}
\end{equation}

If $\mtx{A}$ is rank-deficient, then it has a non-empty null space and the solution to Equation~\eqref{mlew} or \eqref{mle} is not unique. 
But the inversion in Equation~\eqref{mlewsol} or~\eqref{mlesol} can still be performed in the orthogonal complement of this null space. 
For most purposes, this amounts to choosing the solution of minimum norm, \ie the one with no component in the null space of ${\mtx{A}}$.

Although $\mtx{A}$ is sparse, its size makes explicit inversion of $[\transp{\mtx{A}}\mtx{A}]$ intractable but solutions can be computed via an iterative scheme, like the Lanczos bi-diagonalization named ``LSQR'' \citep{Paige82}, which converges to the minimum norm solution when $\mtx{A}$ is singular. 

It is well known that SRT is an ill-posed or under-constrained problem resulting from the limited information available and further complicated by specific problems associated with coronal images obtained with externally-occulted coronagraphs.
\begin{enumerate}
 \item The occulter blocks the solar disk and partially the inner corona (depending upon the exact design of the coronagraph); hence, the images contain information only beyond some minimum projected radius, for instance $\approx$ \SI{2.2}{\Rsun} in the case of LASCO-C2, resulting in so-called hollow projections.
 \item An additional difficulty comes from the outer limit $r_{max}$ of the \fov{} of the instrument whereas the LOS integrals include contributions from electrons located beyond this distance.
It is necessary to include in the reconstruction all parts of the corona that contribute to the observed signal and that includes those beyond $r_{max}$, see the discussion in \citet{Frazin10}.
 \item The image cadence is usually limited to a few images per day while the Sun and its corona rotates at about \SI{13}{\degree} per day, thus resulting in limited angular sampling. 
 \item A final minor point concerns the view of the corona restricted to the ecliptic plane implying that each of the line integral paths is parallel to this plane which is not perpendicular to the sun rotation axis.
\end{enumerate}
% which will be discussed later on.
%DV
These problems lead to a measurement matrix ($\mtx{A}$) that is not only singular, but poorly conditioned (small singular values). 
The direct consequence is that the solution of Equation~\eqref{mlewsol} or \eqref{mlesol} is very noisy and not robust --- a small change in the data has a large impact on the solution --- because of the division by the very small singular values. 
To circumvent these limitations, we introduce a-priori knowledge of the solution, a process known as regularization.

\subsection{Regularization of the SRT Problem}
%---------------------------------------------

\subsubsection{Tikhonov Regularization}
\label{tikhonov_regul}
%---------------------------------------------

We first introduce the Tikhonov regularization which adds a penalty term to the likelihood function enforcing a-priori properties of the solution, for instance smoothness.
This translates to solving the following equation:
\begin{equation}
 \bvec{\hat{x}} = \argminover{\bvec{x}} 
\norm{\bvec{y}-\mtx{A}\bvec{x}}^2_2 + \lambda^2\norm{\mtx{S}\bvec{x}}^2_2 \,,
\label{mlereg}
\end{equation}
where $\lVert\mtx{S}\bvec{x}\rVert^2_2$ is the regularization term, $\lambda$ the regularization parameter, and $\mtx{S}$ the regularization operator.
The choice of the proper regularization has been thoroughly discussed by \citet{Frazin07} and we follow their recommendations of using the second derivatives with respect to the angular spherical coordinates $(\theta,\phi)$ --- except that we add the distance $r$ in order to remove radial noise --- and a single regularization parameter. 

Equation~\eqref{mlereg} can now be rewritten
\begin{equation}
\bvec{\hat{x}} = \argminover{\bvec{x}} 
\left\lVert 
\begin{pmatrix}
\bvec{y}\\
\bvec{0}
\end{pmatrix}
- 
\begin{pmatrix}
\mtx{A}\\
\lambda \mtx{S}
\end{pmatrix}
\bvec{x} \right\rVert^2_2 \,,
\label{mleregeqstacked}
\end{equation}
whose analytic solution is
\begin{equation}
\bvec{\hat{x}} = (\transp{\mtx{A}}\mtx{A} + \lambda^2\transp{\mtx{S}}\mtx{S})^{-1} \transp{\mtx{A}}\bvec{y} \,,
\label{mleregeqstacked_sol}
\end{equation}
which can be solved in the same manner as the unregularized linear least-square problem.

Having settled the regularization procedure, it remains to properly tune the regularization parameter ($\lambda$) controlling the balance between the fidelity of the solution to the data and the level of regularization. 
The strategy in the specific case of coronographic SRT has been extensively discussed by \citet{Frazin00} and \citet{Frazin02}. 
\citet{Barbey13} later argued that this parameter can be determined in an unsupervised way, using a Bayesian formalism and adopting a probability distribution law for it. 
This however leads to prohibitively time-consuming algorithms, as pointed out, for instance, by \citet{Orieux13}. 
Turning therefore to the supervised method, classical approaches are the L-curve \citep{Hansen92,Hansen93}, cross-validation \citetext{\citealp{Stone74,Allen74,Frazin02} and references therein}, and the generalized cross-validation \citep{Golub79} known as ``GCV''.

The cross-validation requires performing several inversions and, at each time, on a different training set built by removing some of the observed pixels and then predicting the removed ones. 
This is repeated for several values of $\lambda$ and finally, the value leading to the best prediction is selected. 
This procedure is very resource-consuming and we thus focus on the two other. 

The principle of the L-curve is to determine the best compromise by looking at the shape of the curve displaying the residual error ($\norm{\bvec{y}-\mtx{A}\bvec{\hat{x}}(\lambda)}$) versus the regularization semi-norm ($\norm{\mtx{S}\bvec{\hat{x}}(\lambda)}$) for each $\lambda$. 
The optimum point of this L-shaped curve is taken at its maximum curvature.

GCV was developed to find the same compromise as cross-validation by simply minimizing the following analytic expression:
\begin{equation}
\argminover{\lambda}
\frac{m\norm{(\mtx{I} - \mtx{K})\bvec{y} }_2^2}
{(\operatorname{trace}(\mtx{I} -\mtx{K}))^2} \,,
\label{gcv}
\end{equation} 
where the operator $\mtx{K}$, often denoted ``hat-matrix'', produces the estimated observation from the data such that $\mtx{I} - \mtx{K}$ is the operator computing the residual:
\[
\mtx{K}=\mtx{A}(\transp{\mtx{A}}\mtx{A} + \lambda^2\transp{\mtx{S}}\mtx{S})^{-1} \transp{\mtx{A}} \,.
\]

The difficulty with this expression lies in computing the denominator. 
This would in principle require performing several inversions as in the case of simple cross-validation. 
But following \citet{Golub97a}, we can use an approximation limiting the task to performing an inversion on one random vector ($\bvec{u}$) containing -1 or +1 elements with uniform probability.
Then
\begin{equation}
\bvec{\hat{t}}(\lambda) = \transp{\bvec{u}}(\mtx{I} - \mtx{K})\bvec{u}
\label{gcv-trace}
\end{equation}
is an unbiased estimator of $\bvec{t}(\lambda)=\operatorname{trace}(\mtx{I}-\mtx{K})$.  

The regularization introduced in this section should be able to partly cope with the difficulties listed at the end of Section~\ref{formulation}, particularly the low cadence of images. 
In the practical case of the LASCO-C2 images, \citet{Frazin07} studied a two-week time period during which the instrument took about 85 $pB$ images in order to ascertain the importance of the cadence for the quality of the reconstruction. 
Their principal finding is that, beyond a certain cadence, there is no benefit in taking more images, although higher cadences do not have negative consequences. 
This is due to the dynamical nature of the solar corona which imposes a fundamental limitation to the spatial resolution of the reconstruction.

\subsubsection{Constraint on the Lower Bound}
\label{sec_min_bkg}
%---------------------------------------------

As the electron density is a strictly positive function, a positivity constraint can be applied.
The algorithms usually chosen to solve non-negative linear least-square problems are ``NNLS'' \citep{Lawson95} and its modified versions such as ``fast-NNLS'' \citep{Bro97} or the more general ``BVLS'' \citep{Stark95} algorithm which handles arbitrary lower and upper bounds. 
However, the ``L-BFGS-B'' algorithm \citep{Byrd95} developed for the minimization of more general convex functions instead of specific linear least-square minimizations  turns out to be much faster for the same level of accuracy in its latest version \citep{Morales11} and was therefore adopted here.

In practice, the above positivity constraint unfortunately tends to set negative values to zero and is therefore of limited interest as it will be shown later (Section~\ref{sec_stat_dyn}). 
This difficulty may be circumvented by introducing an a-priori electron density minimal background and constraining the reconstructed solution to be larger. 
This is easy to achieve since the L-BFGS-B algorithm is actually able to deal with any box-constraint and we can set the lower bound of the solution to be the minimal background. 
However, we choose another path consisting in subtracting the observed minimal background ($\mtx{A}\bvec{b}$) from the data, then performing the inversion using the positivity constraint, and finally adding the background ($\bvec{b}$) to the solution:
\begin{equation}
 \bvec{\hat{x}} = \bvec{b} + \argminover{\bvec{s}\geqslant 0} 
 \left(  
 \left\lVert \bvec{y}-\mtx{A}\bvec{b} -      \mtx{A}\bvec{s} \right\rVert^2_2 
 +\lambda^2 \left\lVert \mtx{S}\bvec{s} \right\rVert^2_2
 \right)
\,.
\label{eq_stc_back}
\end{equation}
The advantage of this approach versus the direct minimum background lower bound constraint is that the regularization operator is applied to the structures of the corona without the background, since it has been subtracted.

\subsection{Illustration of SRT Reconstruction of an MHD Model of the Corona}
%-------------------------------------------------------------------------

We now make use of the three-dimensional magnetohydrodynamic (MHD) simulation of the solar corona presented in \ref{mhdmodel} to test the above SRT procedure.
The MHD simulation is a steady-state model and we obtain temporal variability with a sequence of steady-state models, each with a different synoptic magnetogram specifying the lower boundary.
The synthetic daily $pB$ images calculated from the 3-D electron density resulting from the MHD simulation are ingested by the computer program and the solutions for the electron density are compared to the original input.
We adopt boundary radial distances on the synthetic images of  $r_0=\SI{2.5}{\Rsun}$ and $r_{max}=\SI{6.2}{\Rsun}$ consistent with the LASCO-C2 images to be analyzed later on.
We perform the reconstruction on an enlarged spherical grid of \SI{8.5}{\Rsun} to include LOS contributions beyond $r_{max}$ as pointed out in Section~\ref{formulation}.
The grid has constant resolutions of \SI{0.1}{\Rsun} in the radial direction and of \SI{3}{\degree} in both longitude and latitude.

\subsubsection{Static Reconstruction of a Static Model Corona}
%------------------------------------------------------------
\label{static_static}

\begin{figure*} % 2
\begin{center}
 \includegraphics[width=\textwidth]{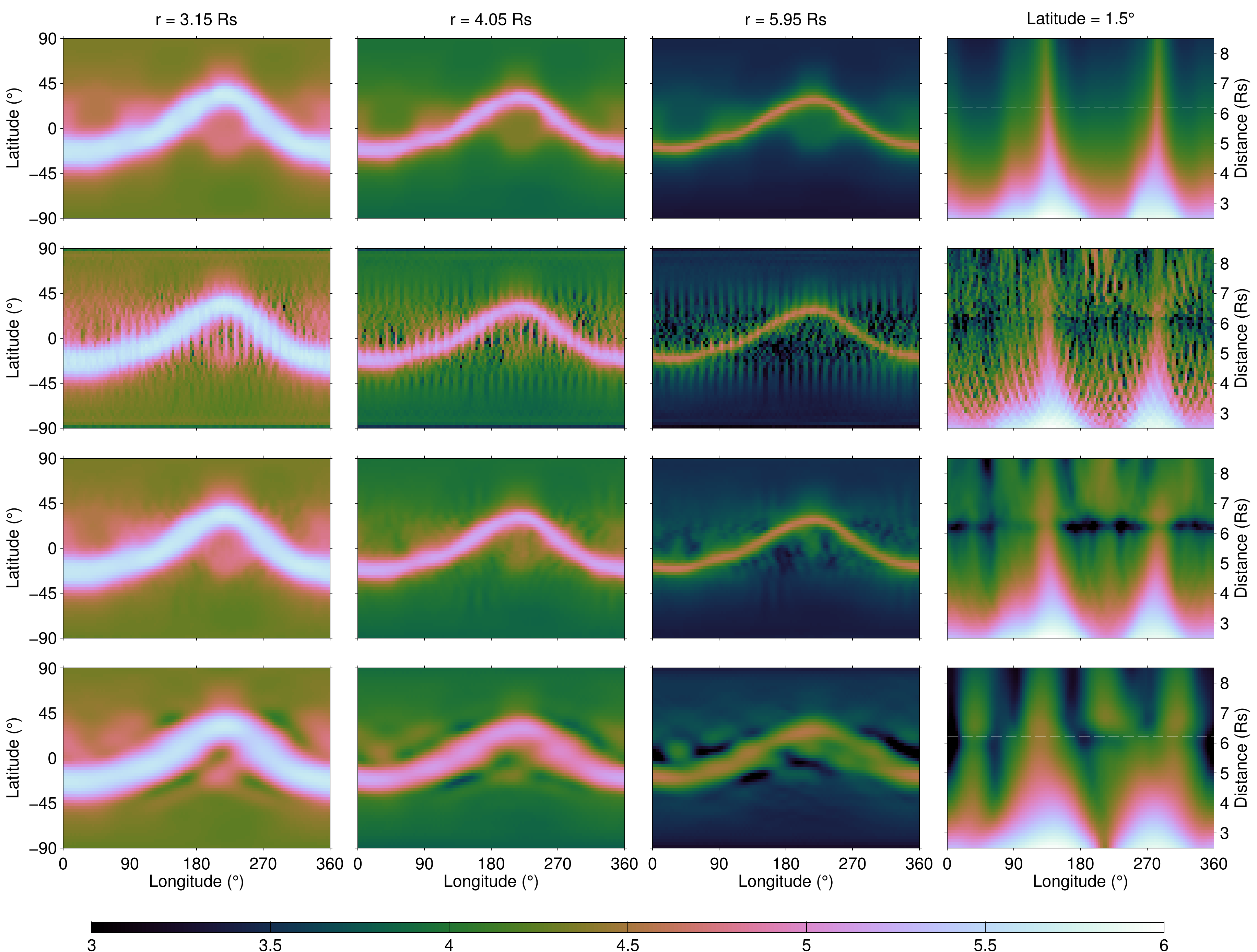}
\caption{Illustration of spatially regularized static reconstructions of a static model for 3 different choices of the regularization parameter.
The 16 panels represent slices of the electron density in units of \si{\cmc} using a logarithmic scale.
The 3 leftmost columns correspond to spherical shells centered at radii of \SIlist{3.15;4.05;5.95}{\Rsun} and the rightmost column corresponds to the equatorial plane (seen in polar coordinates).
The first (top) row corresponds to the model and the next 3 rows to reconstructions with different values of the regularization parameter ($\lambda$): \num{1e-7} (second row), the optimal value \num{1E-6} (third row), and \num{2E-5} (fourth, bottom row).
The white dashed line in the four panels of the rightmost column corresponds to the practical limit of the useful \fov{} (\SI{6.2}{\Rsun}).}
\label{SolutionStat_ModelStat}
\end{center}
\end{figure*}

As a first exercise, we process synthetic images obtained from 14 viewpoints (one per day) as seen from the SOHO orbit using a single steady-state MHD model, that is considering a static corona.
The third row of Figure~\ref{SolutionStat_ModelStat} displays the result for which the regularization parameter ($\lambda$) is determined using the ``GCV'' method.

The excellent agreement with the original MHD electron density distribution proves that the regularization is working very well and is indeed able to cope with the limitations of the coronagraph images as listed at the end of Section~\ref{formulation}. 

In order to further check that the determination of the regularization parameter is optimal, we perform the reconstruction with different values of the regularization parameter ($\lambda$).
Two examples are shown in Figure~\ref{SolutionStat_ModelStat}: when $\lambda$ is larger than the nominal value, the solution is too smooth whereas high frequency noise appears when it is smaller.  Also note the zero density artifacts (hereafter ZDA) that arise when $\lambda$\ is too large.
We finally compute the normalized RMS error between the reconstruction and original electron distribution which we classically define at each radial distance from the Sun as the root mean squared error normalized by the standard deviation of the model:

\begin{equation}\label{rms_eq}
 {\cal E}_{\textnormal{rms}}(r) = \frac{1}{\sigma(r)} 
 \sqrt{\frac{1}{N}\sum_{\theta,\phi} (\hat{x}_{r,\theta,\phi} - x_{r,\theta,\phi})^2}\,,
%\norm{\bvec{\hat{x}} - \bvec{x}}_2
\end{equation}
where 
$$ \sigma(r)^2=\frac{1}{N}\sum_{\theta,\phi} (x_{r,\theta,\phi} - \bar{x}(r))^2$$ 
and 
$$\bar{x}(r)=\frac{1}{N}\sum_{\theta,\phi} x_{r,\theta,\phi} \,.$$

We verify that the chosen optimal value for $\lambda$ is that which minimizes this normalized RMS error.

\subsubsection{Static Reconstruction of a Dynamic Model}
%-------------------------------------------------------
\label{sec_stat_dyn}

\begin{figure*} % 3
(a)\begin{center}
 \includegraphics[width=\textwidth]{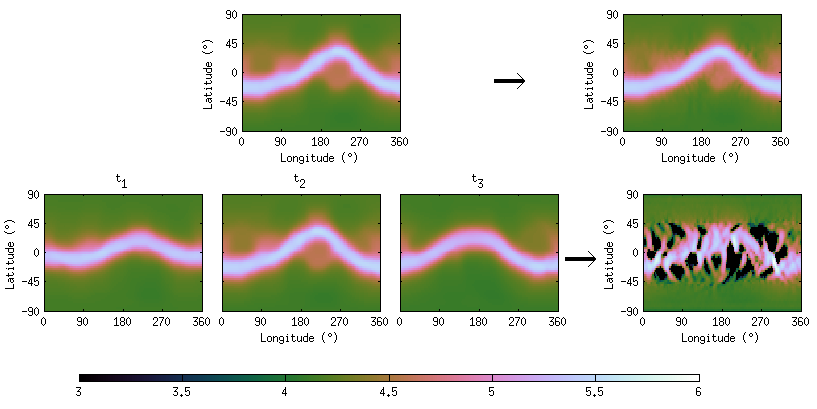}
\end{center}
(b)\begin{center}
 \includegraphics[width=\textwidth]{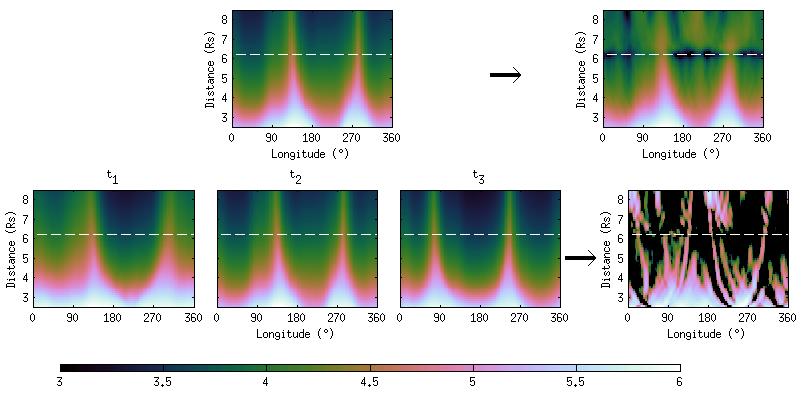}
\end{center}
\caption{(a) Comparison of the standard static reconstruction with spatial regularization applied to two cases of a model corona, a static one (top row) and a dynamic one (bottom row). 
Each image represents a slice at constant radius ($r=\SI{3.45}{\Rsun}$) of the electron density in units of \si{\cmc} using a logarithmic scale. 
The images on the left correspond to the model, and the rightmost image corresponds to the reconstruction.
The time interval between each image of the dynamic model is 6 days.
(b) Same as (a) with the images showing the equatorial plane in polar coordinates.
}
\label{SolutionStat_ModelDyn}
\end{figure*}

We now consider the more realistic case of a dynamic corona. 
For each of the above 14 viewpoints, we input the synthetic images from the dynamic MHD model generated at the actual times LASCO took $pB$ sequences (November 2008).  
Thus, the observation geometry was identical to the case of static reconstruction of a static model corona in Section~\ref{static_static}. 
Figure~\ref{SolutionStat_ModelDyn} compares the two reconstructions for a static and a dynamic corona.
In the latter case, the static reconstruction is unable to cope even with small temporal variations typical of a minimum of solar activity: the ZDAs are overwhelming and render the solution useless.
The problem most likely stems from the slow motion of sharp coronal structures (prominently the streamer belt), as static reconstruction methods cannot distinguish between the effects of rotation and dynamics.

\subsection{Additional Difficulties Associated with SRT}
%-------------------------------------------------------

In addition to the difficulties associated with the facts that the SRT is an ill-posed problem and that the corona varies with time, we briefly summarize below additional difficulties that SRT faces.

\begin{enumerate}
 \item Coronal images are affected by various instrumental problems which are difficult to calibrate and correct, for instance imperfect removal of instrumental stray light and uncertainties in the polarization measurements.
\item
 In addition, in the case of the coronal radiance $B$, assumptions are required to perform the separation of the K and F components which may therefore be imperfect. 
 \item The corona is further perturbed by coronal mass ejections (CMEs); affected frames may be removed or masked to alleviate the problem \citep{Frazin00} but this does not eliminate the abrupt reconfigurations of the global coronal structure that sometime take place following the eruption of a CME \citep{Floyd14}.
\end{enumerate} 
Several of the above difficulties are inherent to the corona and its techniques of observations and thus cannot be circumvented.

%==================================================================================
\section{Static Reconstructions Mitigating Artifacts Due to Temporal Variations}
\label{static}
%==================================================================================
%\subsection{Methods Mitigating the temporal variation}
%------------------------------------------------------

As temporal variations are strongest in the inner co\-rona and tend to decrease with distance, we suspect that the current SRT reconstruction algorithms unduly propagate outward artifacts created by the highly dynamic inner co\-rona, due to the coupled nature of the linear system of equations.
We suggest two different procedures to mitigate this effect, both at the expense of losing optimality in the statistic sense of Equation~\eqref{mlereg}:
i) masking the inner corona when reconstructing the outer part and
ii) radially weighting the electron density to counterbalance its steep gradient.

\subsection{Reconstruction with Multiple Masking}
%----------------------------------------------------

The simple scheme of using two or more masks and juxtaposing the resulting solutions unsurprisingly creates radial discontinuities.
To enforce radial continuity, we introduce multiple masking in a recursive way, starting from the mask of largest radius, and using the solution obtained at step $(n)$ in the reconstruction at step $(n+1)$.

\begin{figure} % 4
\begin{center}
 \includegraphics[width=\columnwidth]{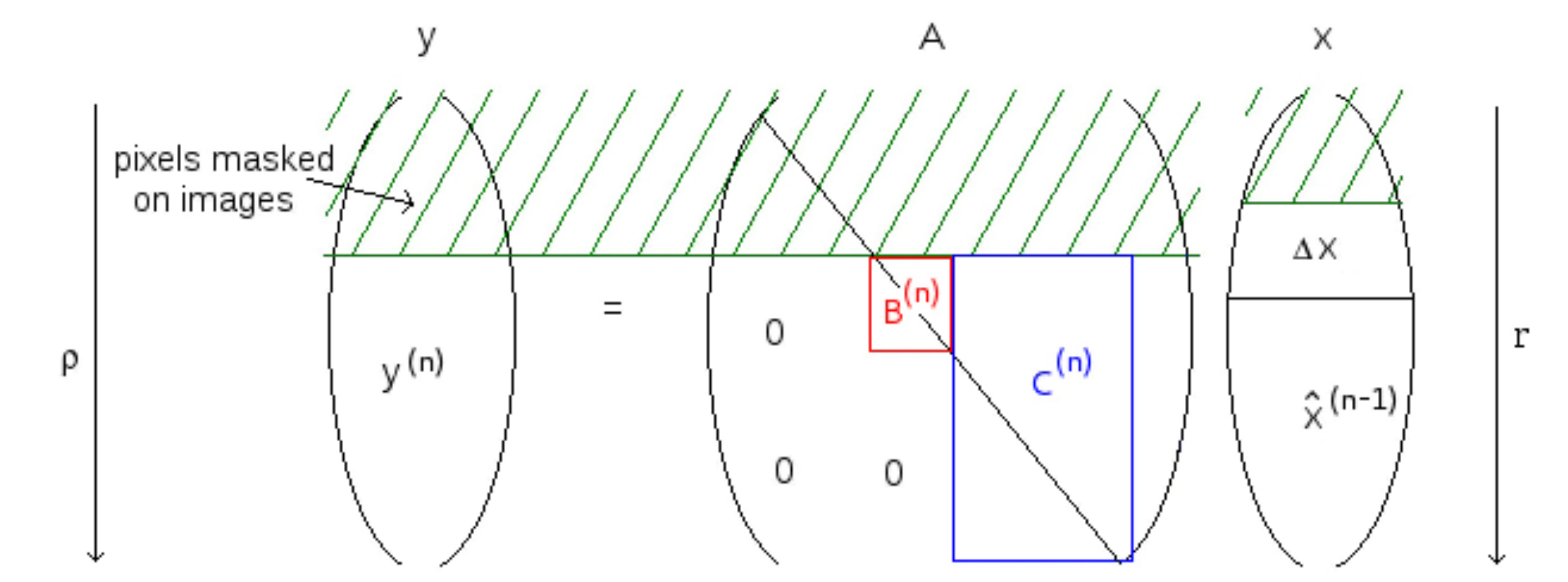}
\caption{Illustration of the sub-matrices used in the multiple masking procedure with successive combination of solutions.}
\label{matppt}
\end{center}
\end{figure}

Let $\{r^{(n)}, n \in [1,\ldots,N]\}$ be the set of decreasing radii of the masks at each step $(n)$ and let $\bvec{\hat{x}}^{(n-1)}$ be the estimated solution at step $(n-1)$ valid from $r^{(n-1)}$ to $r_{\max}$.
At iteration $(n)$, we subtract from the masked data ($\bvec{y}^{(n)}$) the projected image corresponding to $\bvec{\hat{x}}^{(n-1)}$ and solve for the remaining shell of the corona ($\bvec{\Delta x}^{(n)}$) extending from $r^{(n)}$ to $r^{(n-1)}$ (Figure~\ref{matppt}). 
To maintain radial continuity of the regularization operator through each step, we adopt the same scheme and solve the following equation at step~$(n)$:

\begin{equation}
\begin{split}
\bvec{\Delta \hat{x}}^{(n)} = 
 \argminover{\bvec{\Delta x}\geqslant 0} \left( \left\lVert 
\bvec{y}^{(n)}-\mtx{C}^{(n)}\bvec{\hat{x}}^{(n-1)}  - \mtx{B}^{(n)}\bvec{\Delta x}\right\rVert_2^2 \right. \\
 +  \left. \lambda^2 \left\lVert
- \mtx{S}^{(n-1)}\bvec{\hat{x}}^{(n-1)} - \mtx{\Delta S}^{(n)}\bvec{\Delta x} \right\rVert_2^2 \right)
 \,,
\end{split}
\label{mle_mmask_reg}
\end{equation}
where $\mtx{C}^{(n)}$ are the columns going from $r^{(n-1)}$ to $r_{\max}$ of the ``masked'' matrix ($\mtx{A}^{(n)}$) which is the matrix $\mtx{A}$ without the lines corresponding to masked pixels below $r^{(n)}$.
$\mtx{B}^{(n)}$ are the remaining columns going from  $r^{(n)}$ to  $r^{(n-1)}$, see Figure~\ref{matppt}.
$\mtx{S}^{(n-1)}$ is built from the columns of $\mtx{S}$ going from $r^{(n-1)}$ to  $r_{\max}$.
The regularization matrix ($\mtx{\Delta S}^{(n)}$) corresponding to the shell ($\bvec{\Delta x}^{(n)}$) is built from the columns of $\mtx{S}$ going from $r^{(n)}$ to  $r^{(n-1)}$.
 
Note that we could introduce a different value of $\lambda$ at each step. 
We did test this possibility but did not see much change; therefore a unique regularization parameter ($\lambda$) has been used at all steps.

Finally, the estimated solution at iteration $(n)$ is the juxtaposition of the previous solution ($\bvec{\hat{x}}^{(n-1)}$) and the above shell ($\bvec{\Delta x}^{(n)}$):

\begin{equation}
\bvec{\hat{x}}^{(n)} = [\bvec{\Delta x}^{(n)}; \bvec{\hat{x}}^{(n-1)} ] \, , 
\end{equation}
where the semicolon indicates column stacking; thus, $\bvec{\hat{x}}^{(n)}$\ is a column vector consisting of the two smaller column vectors $\bvec{\Delta x}^{(n)}$\ and $\bvec{\hat{x}}^{(n-1)}$.

\subsection{Reconstruction with Radial Weighting}
%---------------------------------------------------
\begin{tolerant}{2000}
Introducing a radial weighting of the images in the least-square minimization requires defining a proper weighting function.
In treating the temporal variations of the corona as noise in the observed images, we note that its variance decreases with increasing distance from the Sun. 
Therefore the coefficients of the diagonal covariance matrix ($\mtx{N}$), which were originally set to the constant $\sigma^2$, can be replaced by a radial function ($f(\bvec{r})$) which expresses an appropriate weighting. 
Writing  $\mtx{N} = \diag(f(\bvec{r}))$ and introducing 
$\mtx{A'}=\diag(\sqrt{f(\bvec{r})})\mtx{A}$ and $\bvec{y'}=\diag(\sqrt{f(\bvec{r})})\bvec{y}$, the equation to be solved is identical to the original Equation~\eqref{mlesol} after substituting 
$\mtx{A}$ by $\mtx{A'}$ and $\bvec{y}$ by $\bvec{y'}$. 
\end{tolerant}

We tested two functions $f(\bvec{r})$: the mean radial profile of the coronal radiance and an exponential function defined by:
\begin{equation}
f(r) = 
\begin{cases}
 e^{\frac{b-r}{c}} & \textnormal{ for } r \leq b\\
 1              & \textnormal{ for } r>b
\end{cases}
\label{wprofile}
\end{equation}
where $b$ is the radius beyond which the weighting is no longer necessary and $c$ controls the sharpness of the weighting for $r_{\min} < r < b$. 
%% note for section application b=5 c=.25

\subsection{Illustration and Comparison of the Methods of Reconstruction}
%------------------------------------------------------------------------

\begin{figure*} % 5
\begin{center}
 \includegraphics[width=\textwidth]{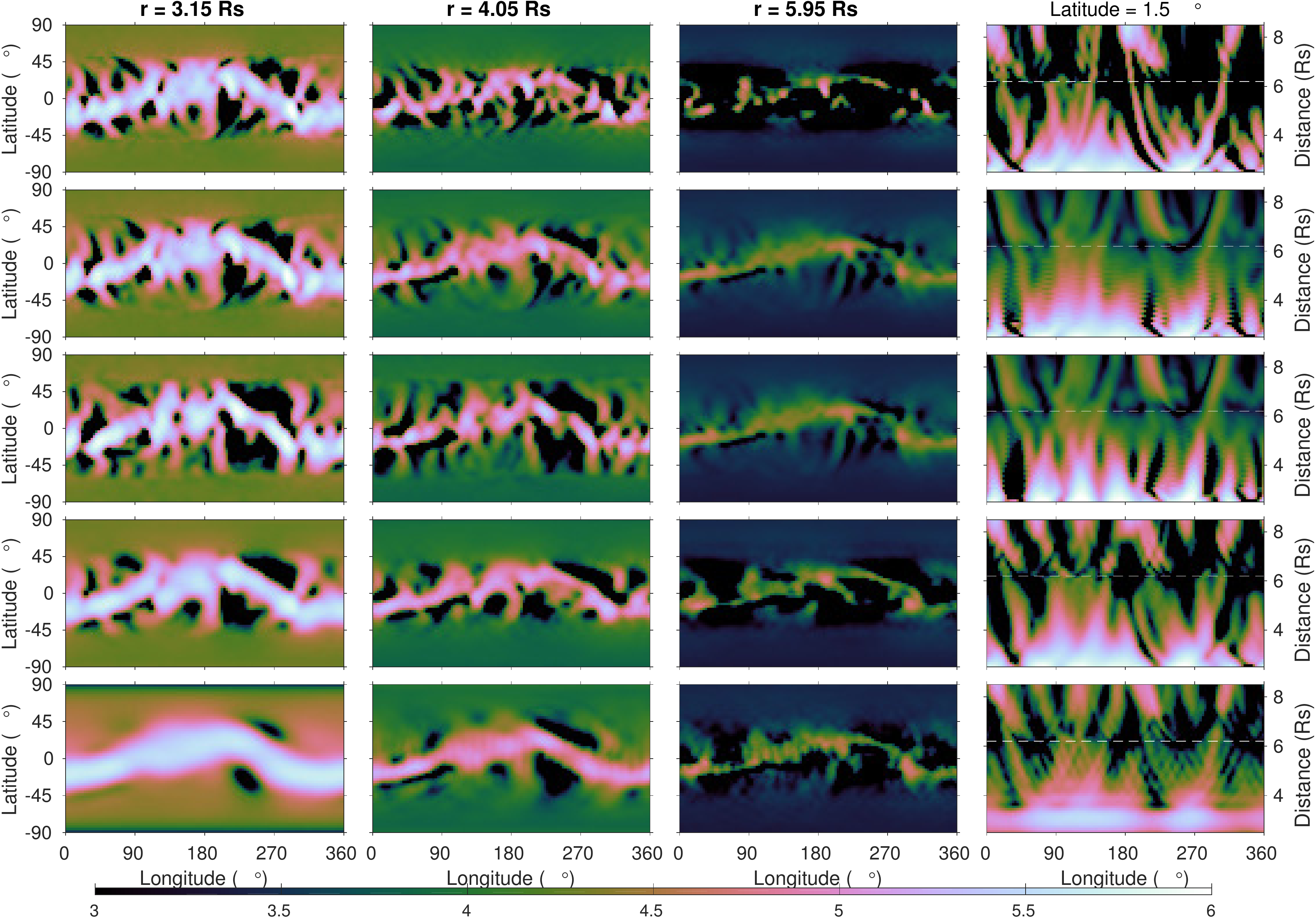}
\caption{Comparison of five static reconstruction procedures applied to a dynamic (MHD) model of the corona as displayed in Figure~\ref{modelAnex}.
The 20 panels represent slices of the electron density in units of \si{\cmc} using a logarithmic scale.
Each row corresponds to one of the static reconstruction methods: the standard spatial regularization (top row), multiple masking with simple juxtaposition (second row) and recursive combination of solutions  (third row), radial weighting with a mean radial intensity profile (fourth row) and an exponential profile with $b=5$ and $c=0.25$ (bottom row).
The 3 leftmost columns correspond to spherical shells at \SIlist{3.15;4.05;5.95}{\Rsun} and the rightmost column to the equatorial plane in polar coordinates.
}
\label{SolutionStat_5methods}
\end{center}
\end{figure*}

\begin{figure*} % 6
\begin{center}
 \includegraphics[width=\textwidth]{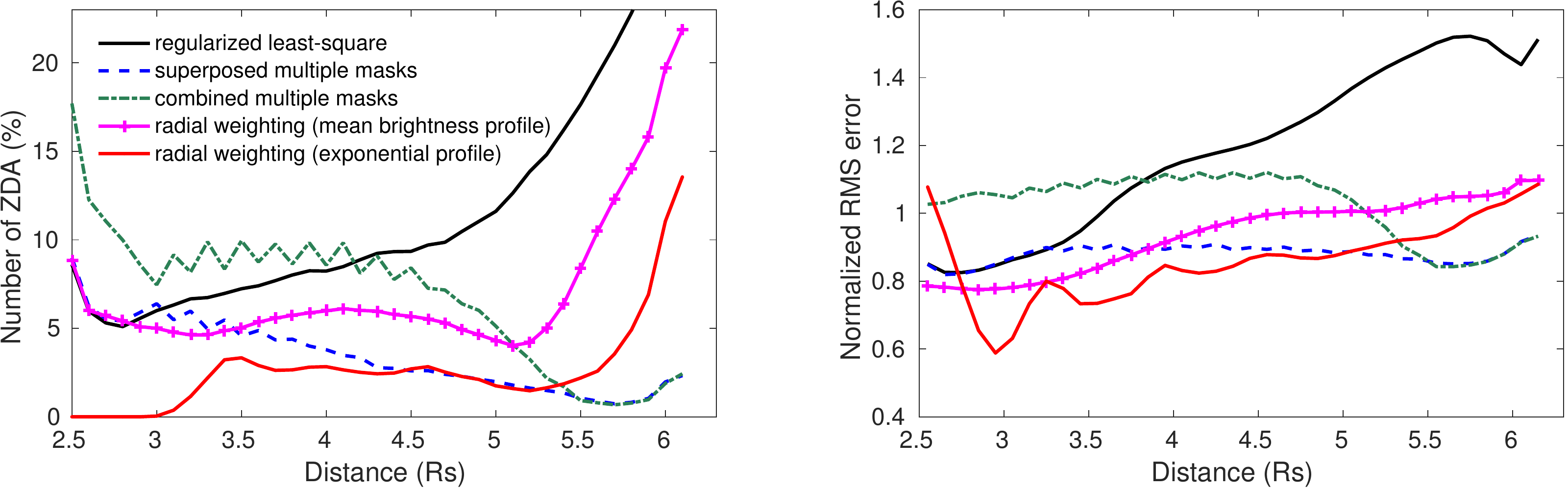}
\caption{Illustrations of the errors associated with the five static reconstruction procedures applied to a dynamic model of the corona, in spherical shells of increasing distance from the Sun.
The left panel presents the percentage of ZDA (Zero Density Artifacts). 
The right panel presents the RMS error normalized by the standard deviation of the model.}
\label{MHD_StatZerosRMS}
\end{center}
\end{figure*}

We first test the above procedures using the MHD model presented in \ref{mhdmodel}.
The results are illustrated and compared in Figure~\ref{SolutionStat_5methods} which displays images of the electron density and in Figure~\ref{MHD_StatZerosRMS} which quantifies the errors associated with each procedure in spherical shells of increasing distance from the Sun:
i) the percentage of ZDAs in the solution (which would have corresponded to negative values in the general case without the positivity constraint) and ii) the RMS error normalized by the standard deviation (see Equation~\eqref{rms_eq}).

Multiple masking, either by juxtaposition or recursive combination, offers an improvement of the reconstruction limited to the outer corona ($> \SI{4}{\Rsun}$) as expected since it avoids propagating dynamic changes to this region. 
Simple juxtaposition naturally introduces radial discontinuities which are alleviated by recursive combination at the price of a degraded reconstruction in the inner part ($\leq \SI{4}{\Rsun}$) to the point of being worse than the standard, unmasked reconstruction. 
This is because of the accumulation of errors from the outer corona to the inner corona during the recursive process.

Radial weighting leads to the opposite trend as the reconstruction is improved in the inner corona but degraded in the outer part, but still superior to the standard one.
Weighting with an exponential profile globally leads to a better reconstruction than weighting with the radiance profile of the corona except inside $\approx \SI{3}{\Rsun}$ where the solution is over-smoothed; indeed it introduces too much weighting as the regularization term dominates in the minimization. 

Inside \SI{3}{\Rsun}, none of the proposed methods improves the reconstruction.
From \SIrange{3}{5}{\Rsun}, the weighting method with an exponential profile gives the best result whereas beyond \SI{5}{\Rsun}, this is achieved by multiple masking with juxtaposition.
In summary, these methods allow improving the quality of the reconstruction in the regimes where they perform best, decreasing the percentage of ZDA and the RMS error by a factor $\approx2$ compared to the standard method but still much inferior to the performance of the reconstruction of a static corona.

An interesting remark concerns the number of ZDA which is generally a good proxy for the RMS error. 
Therefore, it offers a relevant quality factor for the reconstruction of real data for which the error is obviously not available.

\begin{figure*} % 7
\begin{center}
 \includegraphics[width=\textwidth]{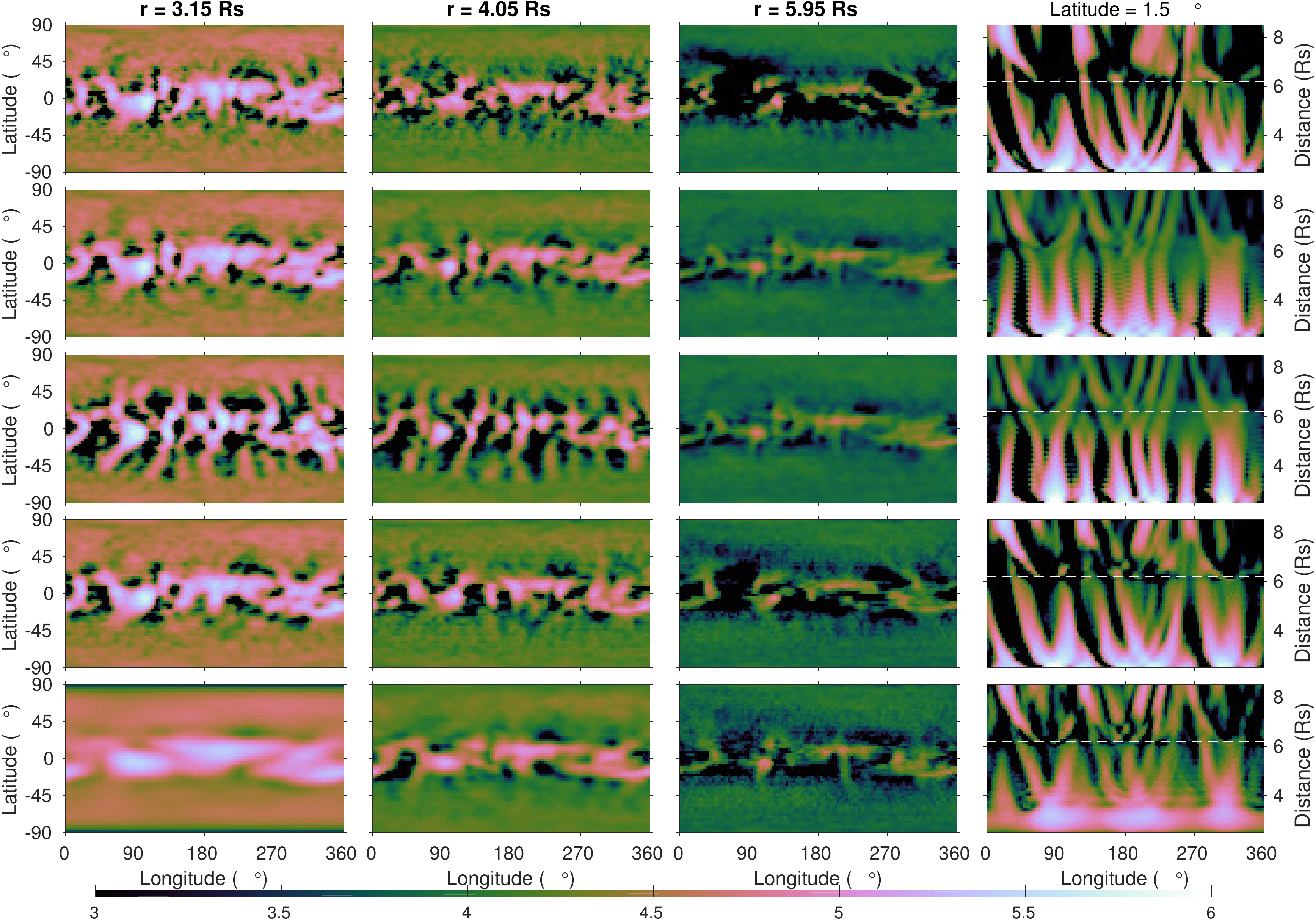}
\caption{Comparison of the five static reconstruction procedures applied to 53 LASCO-C2 $pB$ images. 
The 20 panels represent slices of the tomographic determinations of the electron density in units of \si{\cmc} using a logarithmic scale.
The 3 leftmost columns correspond to spherical shells with radii of \SIlist{3.15;4.05;5.95}{\Rsun} and the rightmost column to the equator.
Each row corresponds to one of the static reconstruction methods: the standard spatial regularization (first row), multiple masking with simple juxtaposition (second row) and successive combination (third row) of solutions, radial weighting with a mean radial intensity profile (fourth row) and an exponential profile (fifth row).}
\label{SolutionStat_5methods_lasco}
\end{center}
\end{figure*}

\begin{figure*} % 8
\begin{center}
 \includegraphics[width=\textwidth]{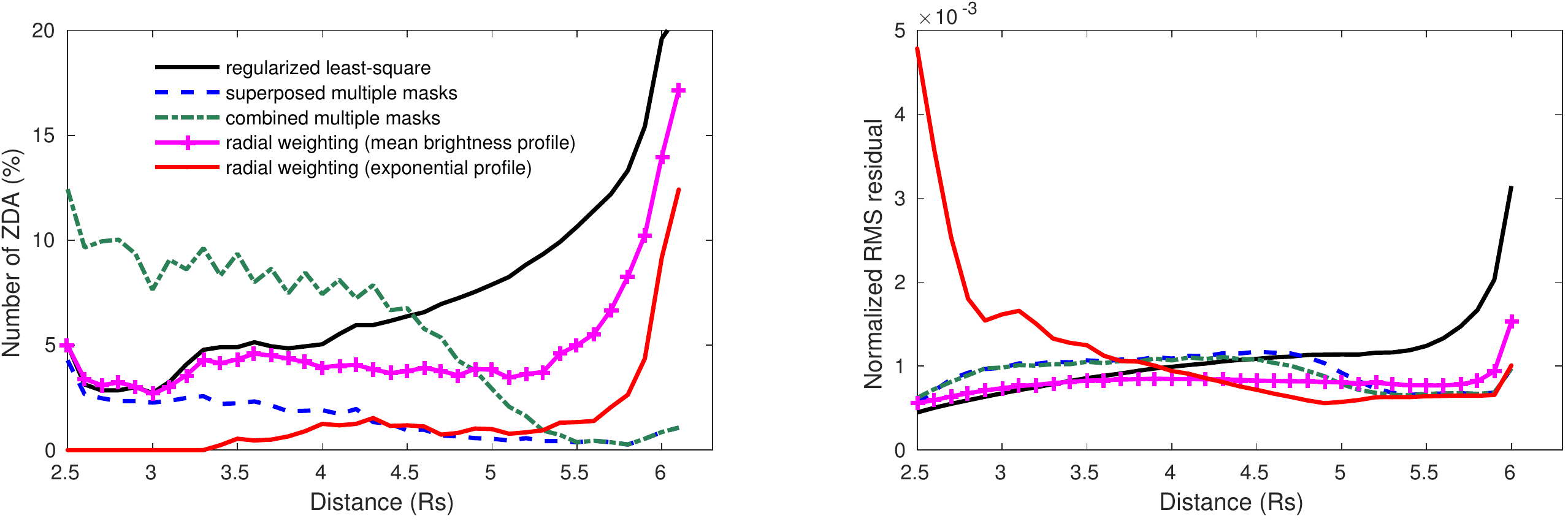}
\caption{Illustrations of the errors associated with the five static reconstruction procedures applied to 53 LASCO-C2 $pB$ images, in spherical shells of increasing distance from the Sun.
The left panel presents the percentage of ZDA. 
The right panel presents the RMS residual error normalized by the standard deviation of the image pixels.}
\label{lasco_stat_zeros}
\end{center}
\end{figure*}

We extend the comparison to the set of 53 LASCO-C2 $pB$ images as described in \ref{lasco} and display the results in Figure~\ref{SolutionStat_5methods_lasco}.
Since we do not have a model, we cannot compute the error of the reconstruction; instead, we compute the residual errors between the observed and the synthetic images calculated from the reconstructed $N_e$ (Figure~\ref{lasco_stat_zeros}).
The pros and cons of the different procedures are consistent with those reached in the case of the MHD model. 
Incidentally, the percentages of ZDA are similar which suggests that the MHD model is adequate to perform the validity tests.

%==================================================================================
\section{Time--Dependent Reconstruction}
\label{dynamic}
%==================================================================================

\subsection{General Formalism}
%-----------------------------

As we have seen, time-dependency of the corona is the fundamental difficulty for SRT.  
Strategies for mitigating the problem were first discussed in \citet{Frazin05b}.  
A Kalman filter solution was finally implemented by \citet{Butala10}, but the performance was not greatly superior to static reconstruction, due to the highly under-determined nature of attempting to infer the evolving state of the corona given only one or a few simultaneous viewpoints.
\subsubsection{Spatio-temporal Regularization} 
%--------------------------------------------

Several \spatiotemp{} regularization methods exist to solve the problem \citep{Zhang05}.
We focus our attention to:
\begin{itemize}
  \item the state-space model using Kalman smoother implemented by \citet{Butala10};
  \item the multiple constraint regularization mentioned by \citet{Barbey08} and \citet{Barbey13}.
\end{itemize}
The theoretical solutions of both methods are identical for some specific choices of the operators and covariance matrices: 
i) scalar multiple identity matrices for both the Kalman state transition matrix and the time covariance matrix in the case of the ``Kalman'' approach and 
ii) an upper or lower bi-diagonal time-regularization operator for the multiple constraint approach. 
In this work, we choose the multiple constraint method since it is simpler to implement and offers a direct interpretation of the smoothness constraints in space and time. 

% description of the method
\begin{tolerant}{1000}
The vector ($\bvec{x}$) to be reconstructed is now time-dependent.
Let stack each static corona ($\bvec{x}_i$) at each time ($t_i$) in one single vector: $\bvec{\bar{x}} = \{\bvec{x}_1; \bvec{x}_2; \ldots; \bvec{x}_{Nt} \}$. The \spatiotemp{} linear measurement system becomes:
\begin{equation} 
\bvec{\bar{y}} = \mtx{\bar{A}}\bvec{\bar{x}} \,,
\end{equation}
where $\bvec{\bar{y}}=\bvec{y}=\{\bvec{y}_1; \bvec{y}_2; \ldots; \bvec{y}_{Nt} \}$ are the stacked vectors of pixels radiance in each image. This vector represents the observed data.
It has the same number of elements and is arranged in the same way as the $\bvec{{y}}$ vector of the static reconstruction. 
The dynamic measurement matrix ($\mtx{\bar{A}}$) is composed of the blocks of the static matrix ($\mtx{A}$) that correspond to the measurements at a given observation time. If

\begin{equation}
\mtx{A}=
\begin{pmatrix}
\mtx{A}_1\\
\mtx{A}_2\\
\vdots\\
\mtx{A}_{N_t}
\end{pmatrix}
\,\textnormal{, then }
\mtx{\bar{A}}=
\begin{pmatrix}
\mtx{A}_1 &          &        &       \\
          &\mtx{A}_2 &        &       \\
          &          & \ddots &       \\
          &          &        & \mtx{A}_{N_t}
\end{pmatrix} \,.
\end{equation}
\end{tolerant}

We then restrict our procedure to the simplest form of the multiple constraint regularization with one ``pure'' spatial and one ``pure'' temporal regularization where the spatial (resp.\ temporal) constraint is the same at each time instant (resp.\ voxel).

The solution is then:
\begin{equation}
 \bvec{\hat{\bar{x}}} = \argminover{\bvec{\bar{x}}\geqslant 0} \norm{\bvec{y}-\mtx{\bar{A}}\bvec{\bar{x}}}^2_2 + \lambda_s^2\norm{\mtx{\bar{S}}\bvec{\bar{x}}}^2_2 + \lambda_t^2\norm{\mtx{\bar{T}}\bvec{\bar{x}}}^2_2 \,,
\label{dyn_mle}
\end{equation}
where $\mtx{\bar{S}} = \mtx{I}_{Nt}\otimes \mtx{S}$ and $\mtx{\bar{T}}=\mtx{T}\otimes \mtx{I}_{N{\textnormal{voxels}}}$ are the ``pure'' spatial and temporal regularizations, $\lambda_s$ and $\lambda_t$ being the spatial and temporal regularization parameters.
The symbol $\otimes$ represents the Kronecker product and  $\mtx{I}_{Nt}$ (resp. $\mtx{I}_{N{\textnormal{voxels}}}$), the identity matrix of dimension $N_t$, the number of time steps or images (resp.\ the identity matrix of dimension $N_{\textnormal{voxel}}$, the number of voxels). 

Note that $\mtx{\bar{T}}$ applies to the same spatial measurements from all time instants and constrains its temporal behavior according to $\mtx{T}$. 
Each row of $\mtx{T}$ acts as a temporal high-pass filter selectively passing undesirable components to be minimized in the reconstruction.

 For the temporal regularization operator $\mtx{T}$, we choose to implement the first derivative to enforce smoothness. 
This operator is bi-diagonal since it filters two consecutive states for each voxel.
In this case, a Kalman smoothing with an identity state-transition matrix theoretically produces the same solution, but the convergence algorithm is different.

As in Equation~\eqref{mleregeqstacked}, we can stack the observation and regularization matrices in one single matrix and rewrite Equation~\eqref{dyn_mle} as a simple, non-negative linear least-square minimization which can be solved using the L-BFGS-B algorithm:
\begin{equation}
\bvec{\hat{\bar{x}}} = \argminover{\bvec{\bar{x}}\geqslant 0} 
\left\lVert 
\begin{pmatrix}
\bvec{y}\\
\bvec{0}\\
\bvec{0}
\end{pmatrix}
- 
\begin{pmatrix}
\mtx{\bar{A}}\\
\lambda_s \mtx{\bar{S}}\\
\lambda_t \mtx{\bar{T}}
\end{pmatrix}
\bvec{\bar{x}} \right\rVert^2_2 \,.
\label{dyn_mleregeqstacked}
\end{equation}

The joint tuning of the two regularization parameters $\lambda_s$ and $\lambda_t$ becomes now a two-dimensional minimization and the GCV criteria can still be applied. 

Some results illustrating this \spatiotemp{} regularization applied to the MHD model and LASCO images are presented and discussed in Section~\ref{dynrec_model} and~\ref{dynrec_lasco} below.

\subsubsection{Co-rotating Regularization}
%-----------------------------------------

 We introduce another regularization operator, this time acting jointly in the spatial and time domain, to improve the \spatiotemp{} regularization.
Its purpose is to prevent the concentration of the density in the vicinity of the plane of the sky (containing the Sun's center).  This plane rotates in the Carrington coordinate system, as it is always orthogonal to the observer's LOS.
Thus we construct an operator that filters the corona, selecting the undesirable co-rotating components, and hope that such components are prominently artifacts and not intrinsic to the corona itself.
%Note that the longitude-invariant component belongs to this category and should be kept in the solution.
The technical detail of the construction of this regularization operator ($\mtx{C}$)  is presented in \ref{CoRot}.

%final minimization equation
The new minimization problem to be solved is
\begin{equation}
\begin{split}
 \bvec{\hat{\bar{x}}} = \argminover{\bvec{\bar{x}}\geqslant 0} 
 \left(\norm{\bvec{y}-\mtx{\bar{A}}\bvec{\bar{x}}}^2_2 + \lambda_s^2\norm{\mtx{\bar{S}}\bvec{\bar{x}}}^2_2 + \lambda_t^2\norm{\mtx{\bar{T}}\bvec{\bar{x}}}^2_2 \right.\\ 
\left. + \lambda_c^2\norm{\mtx{C}\bvec{\bar{x}}}^2_2 \right)\,.
\end{split}
\label{dyn_corot_mle}
\end{equation}

We now have three regularization parameters; their optimal determination is not straightforward and will be discussed when applied to a model and to LASCO images.

%\subsubsection{Minimal background prior}

\subsection{Application to the Dynamic Model of the Corona}
%-------------------------------------------------------------------

\subsubsection{Reconstruction with Spatial and Temporal Regularizations: Influence of the Relative Weight of the Two Regularizations}
\label{dynrec_model}
%--------------------
\begin{tolerant}{1000}
We now perform time-dependent tomography using Equation~\eqref{dyn_mle} and 14 synthetic $pB$ images from the MHD model of the corona described in \ref{mhdmodel}.
Figure~\ref{DynamicSolution_RsRt_r405} displays reconstructions with several choices of the temporal regularization parameter $\lambda_{t}$, the value of $\lambda_{s}$ being equal to the optimum found in the case of static reconstructions ($\lambda_{s}=\num{2.2E-6}$). 

\begin{figure*} % 9
\begin{center}
 (a)\includegraphics[width=.8\textwidth]{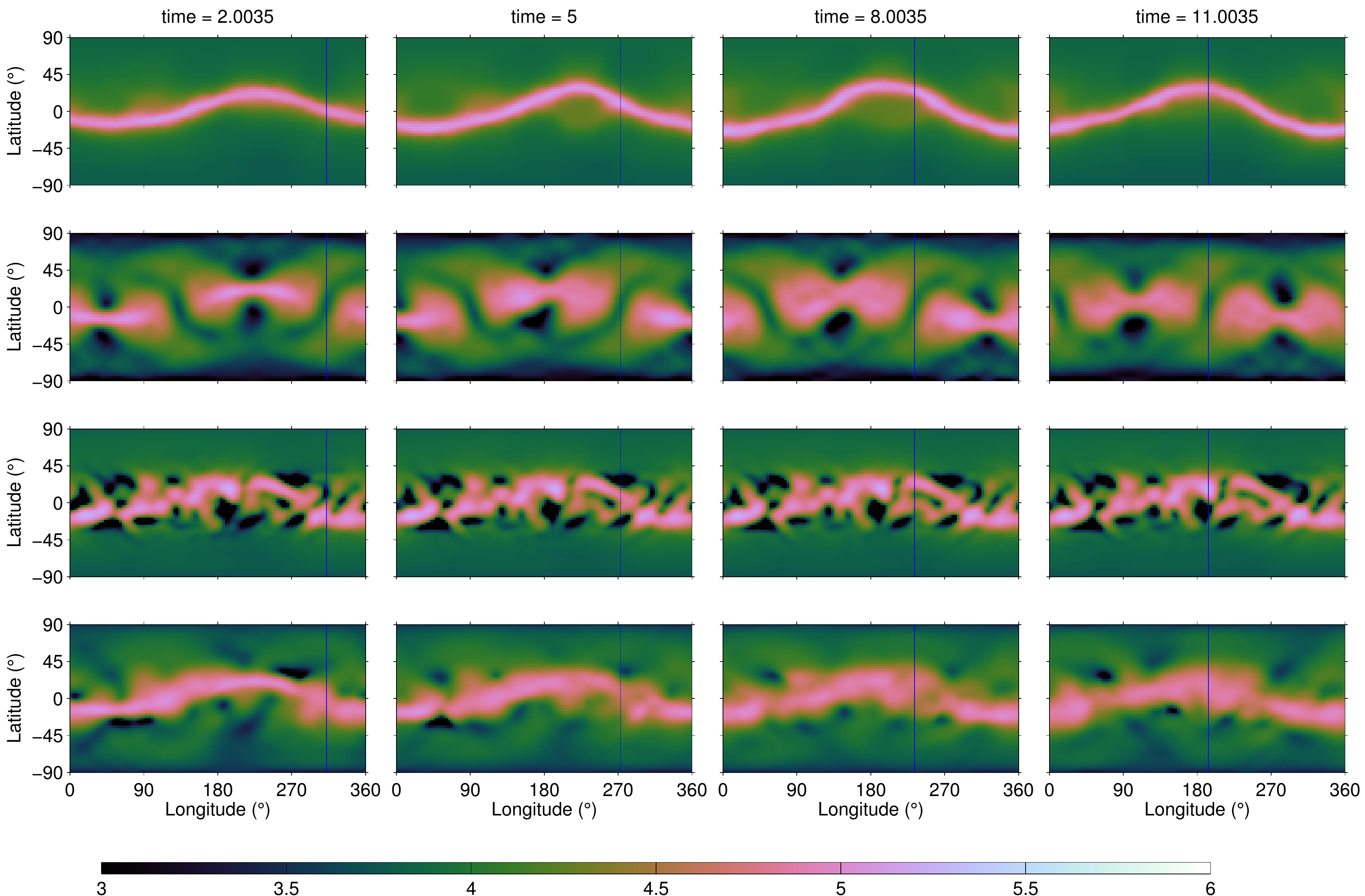}
\end{center}
\begin{center}
 (b)\includegraphics[width=.8\textwidth]{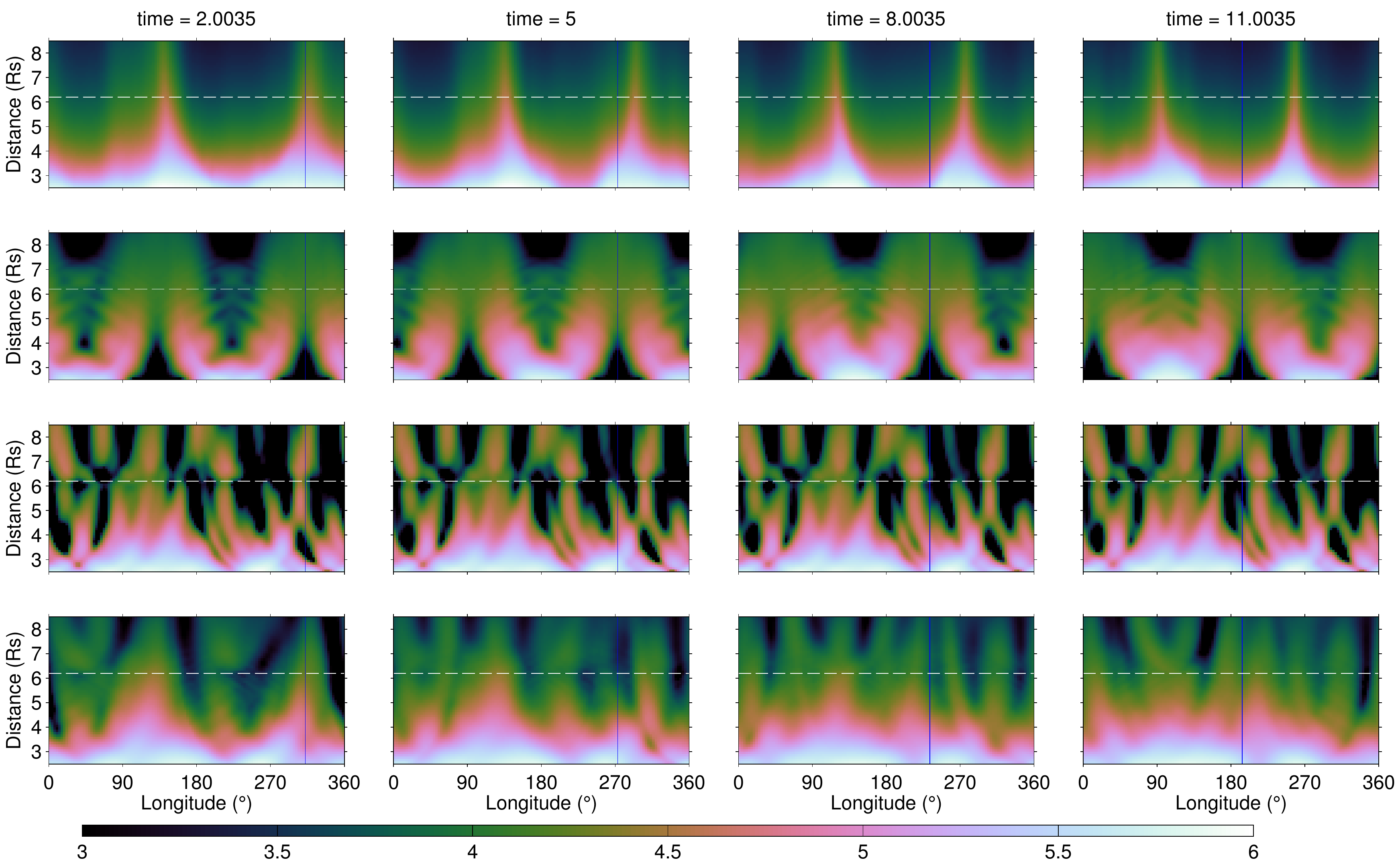}
\end{center}
\caption{(a) Illustration of the \spatiotemp{} regularized dynamic reconstruction of a dynamic model, for 3 different settings of the regularization parameters.
The 16 panels represent shells at a constant radius ($r=\SI{4.05}{\Rsun}$) of the electron density in units of \si{\cmc} using a logarithmic scale.
Each column corresponds to a reconstruction of the corona at a given time.
The first (top) row corresponds to the dynamic model and the next 3 rows to the different reconstructions with the temporal regularization parameter ($\lambda_{t}$) set to 0 (second row), \num{2e-5} (third row), and \num{1.7e-6} (fourth, bottom row).
The value of the spatial regularization parameter is the same for the three reconstructions ($\lambda_{s}=\num{2.2e-6}$).
The blue vertical line corresponds to the Carrington longitude of the observer.
(b) Same as (a)with the panels representing the equatorial plane in polar coordinates.}
\label{DynamicSolution_RsRt_r405}
\end{figure*}
\end{tolerant}

First, looking at the bottom row of this figure corresponding to the optimal choice of the regularization parameters, we note a net improvement over all static reconstructions, even the solutions that use masking or radial weighting (see the second column of Figure~\ref{SolutionStat_5methods}). 
This solution exhibits fewer ZDA  but there are not completely eradicated and we do not reach the quality of the ideal case of reconstructing a static model as shown in Figure~\ref{SolutionStat_ModelStat}. 
A quantitative assessment will be presented in Section~\ref{discomp}.

Second, we find that the solution is highly sensitive to the temporal regularization parameter ($\lambda_t$).
If $\lambda_t$ is large, the time variation of the solution is annihilated and the solution has the same drawback (many ZDA) as the static inversion.
On the flip side, if $\lambda_t$ is low, the time variation of the solution is largely unconstrained and exhibits the same pitfalls as those crippling the time-dependent reconstruction with spatial regularization alone. 
The major problem in this case is a concentration of material in the plane of the sky corresponding to the unique view-point associated with each time/image and corollary, large voids in front of (and behind) the Sun as viewed from the observer. 

\begin{figure*} % 10
\begin{center}
 \includegraphics[width=0.45\textwidth]{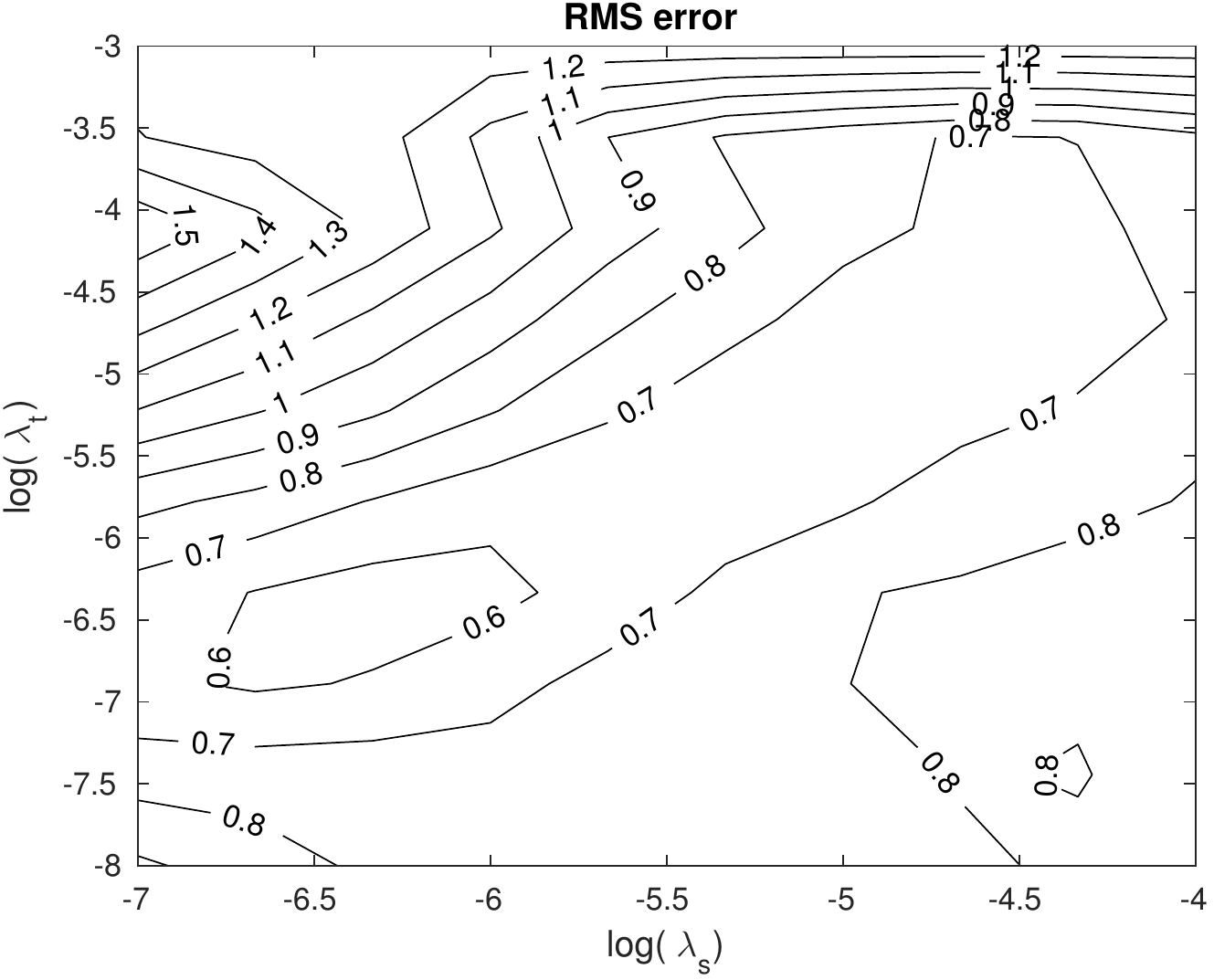}
 \includegraphics[width=0.45\textwidth]{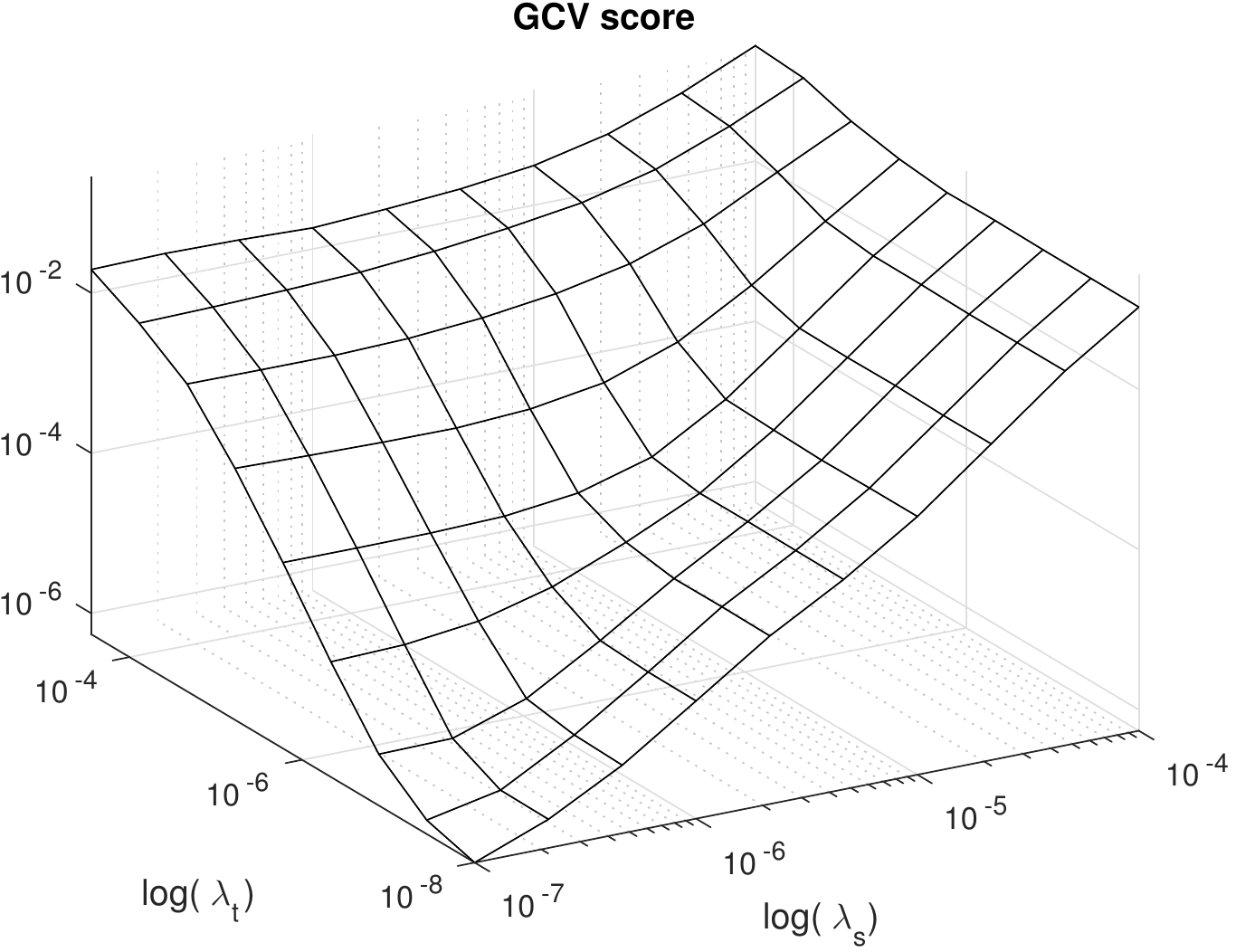}
 \includegraphics[width=0.45\textwidth]{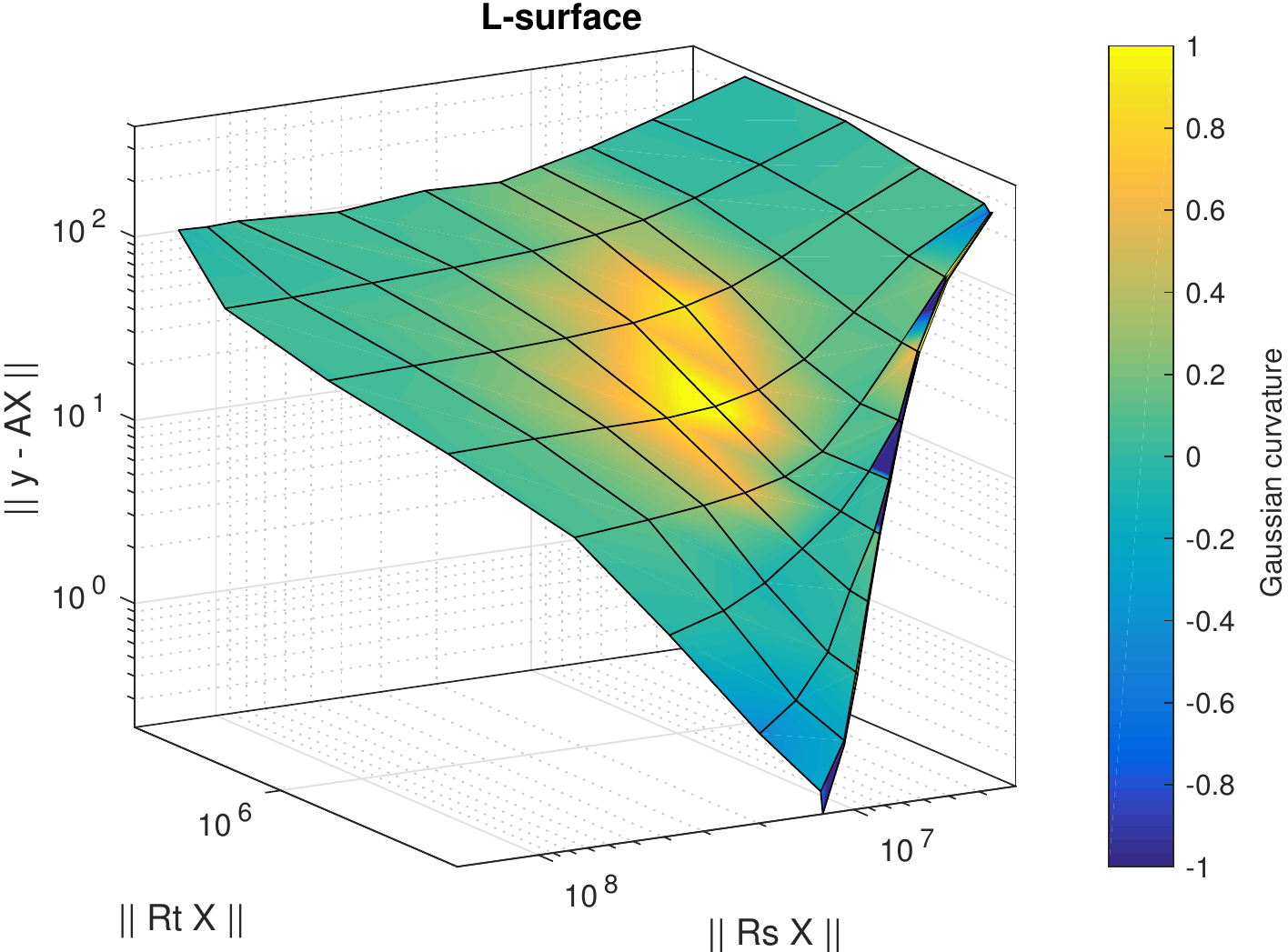}
 \includegraphics[width=0.45\textwidth]{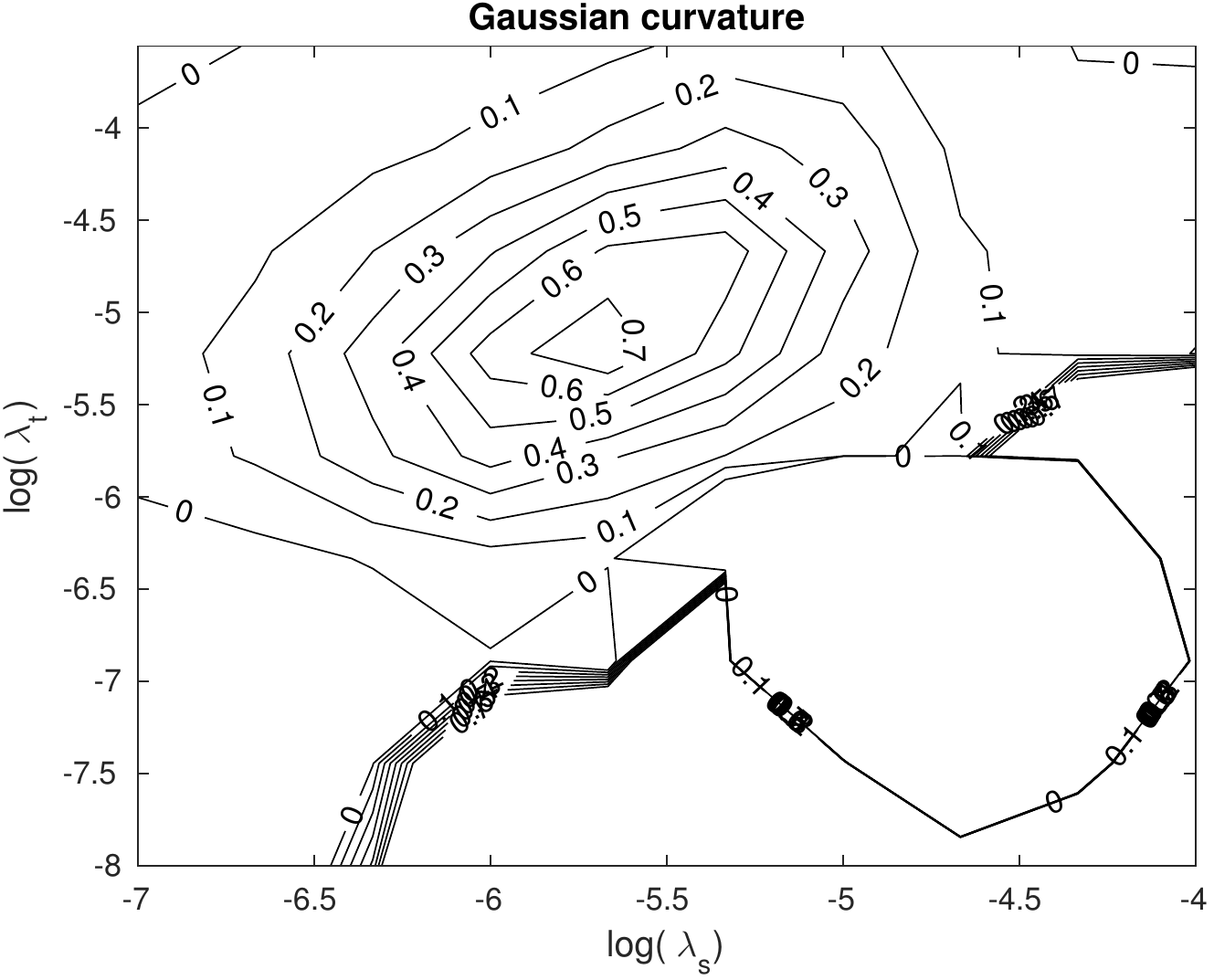}
\caption{Illustration of the RMS reconstruction error (top-left), GCV score (top-right), L-surface (bottom-left), L-surface Gaussian curvature (bottom-right) as a function of the spatial ($\lambda_{s}$) and temporal ($\lambda_{t}$) regularization parameters.}
\label{rms_errors_RsRt}
\end{center}
\end{figure*}

The optimal values of the spatial and temporal regularization parameters are verified by calculating the RMS error normalized by the variance --- see Equation~\eqref{rms_eq} --- averaged over the distance to the Sun and taking into account only 10 days out of 14 since the first and last two days are less regularized in time. 
The result is displayed in the top-left panel of Figure~\ref{rms_errors_RsRt} and shows a sharp transition at $\lambda_{t}$ around \num{3e-4}. 
Above this threshold, the reconstruction swings to a static reconstruction regime with a large number of ZDA. 
The shape of the contours organized along the diagonal indicates a coupling between the two parameters which implies that the optimization is prominently controlled by the ratio of the two parameters. 

The optimal settings of these regularization parameters has to be done without knowing the reconstruction error.
We therefore test the two different scores described at the end of Section~\ref{tikhonov_regul}, namely the CGV and L-curve.
As shown in the top-right panel of Figure~\ref{rms_errors_RsRt}, the GCV score fails since its minimum is only asymptotically reached for $\lambda_{t}=\lambda_{s}=0$.
This behavior is well described by \citet{Hansen92} in the case of correlated errors that is exactly the case here: when the temporal regularization is too weak the solution has a smaller residual but is non-physical, exhibiting concentration of material in the plane of the sky, instead of being simply randomly noisy.
On the other hand, the L-curve method is able to find an optimal setting as shown in the two bottom panels of Figure~\ref{rms_errors_RsRt}.
In the case of two regularization parameters, the L-curve becomes an ``L-surface'' (bottom-left panel) where the optimal setting corresponds to the maximum of the Gaussian curvature (bottom-right panel).
Knowing that the L-curve has a tendency to over-regularize \citep{Golub97a}, we chose optimal values slightly smaller than those given by the L-surface maximum curvature and still along the maximum diagonal of the RMS error plot: $\lambda_s=\num{2.2E-6}$, $\lambda_t=\num{1.7E-6}$.

\subsubsection{Reconstruction with the Co-rotating Regularization}
\label{rec_corot_mhd}
\begin{figure*} % 11
\begin{center}
 \includegraphics[width=\textwidth]{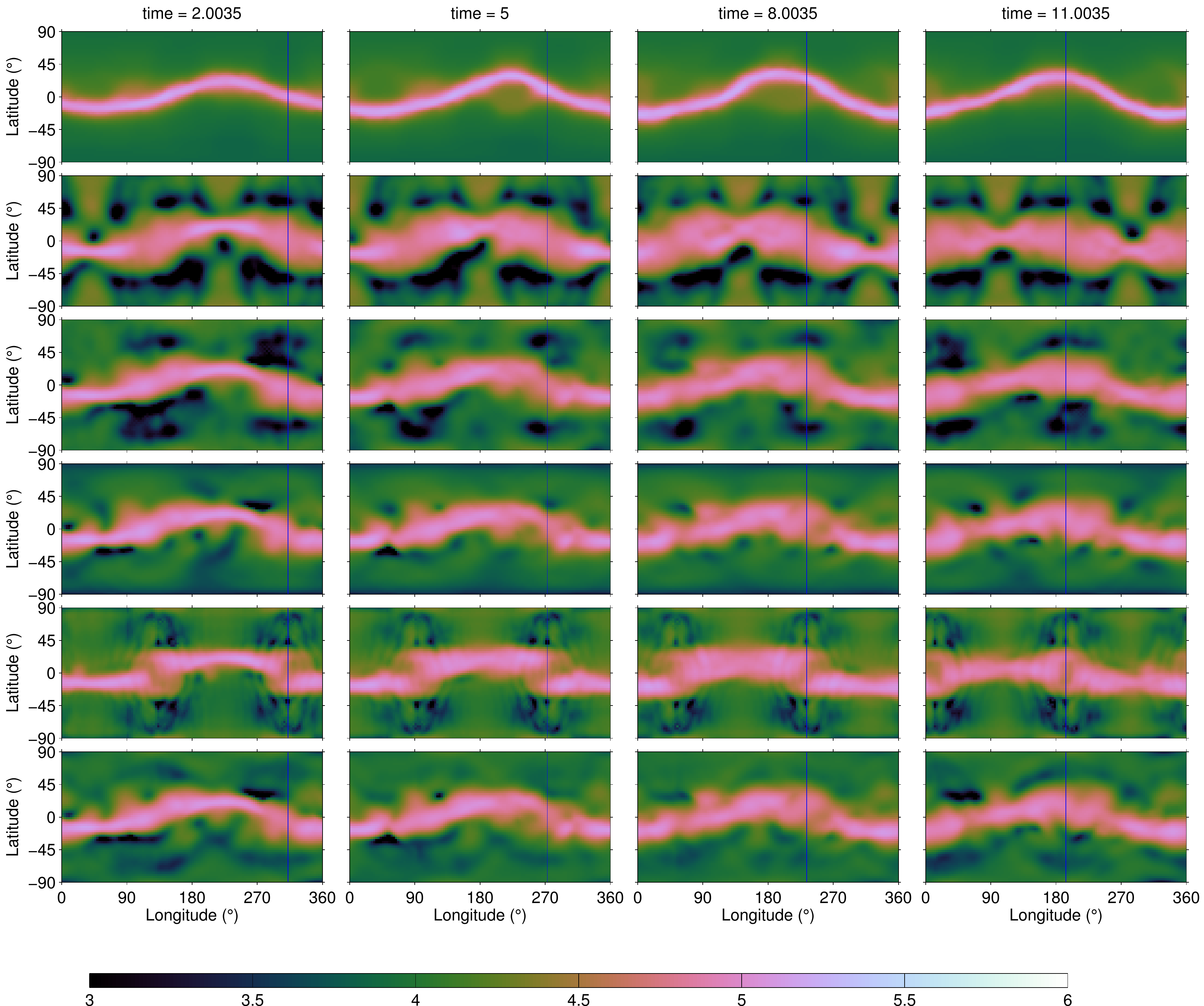}
\caption{Illustration of dynamic reconstructions of a dynamic model using co-rotating regularization.
The 24 panels represent spherical shells at \SI{4.05}{\Rsun} of the electron density in units of \si{\cmc} using a logarithmic scale.
The first row is the dynamic model and the last 5 rows show the reconstructions with various settings of the temporal and \corotating{} regularization and with the same spatial regularization ($\lambda_{s}=\num{2.2e-6}$).
From top to bottom: no temporal regularization and \corotating{} optimum regularization ($\lambda_{t}=0, \lambda_{c}=\num{2e-7}$), small temporal regularization and \corotating{} optimum regularization ($\lambda_{t}=\num{1e-6}, \lambda_{c}=\num{2e-7}$), temporal regularization optimum and small \corotating{} regularization ($\lambda_{t}=\num{1.7e-6}, \lambda_{c}=\num{1e-10}$), temporal regularization optimum and large \corotating{} regularization ($\lambda_{t}=\num{1.7e-6}, \lambda_{c}=\num{1e-5}$), temporal and \corotating{} optimum regularizations ($\lambda_{t}=\num{1.7e-6}, \lambda_{c}=\num{2e-7}$).}
\label{DynamicSolution_RsRtRc_r405}
\end{center}
\end{figure*}

\begin{figure*} % 12
\begin{center}
 \includegraphics[width=\textwidth]{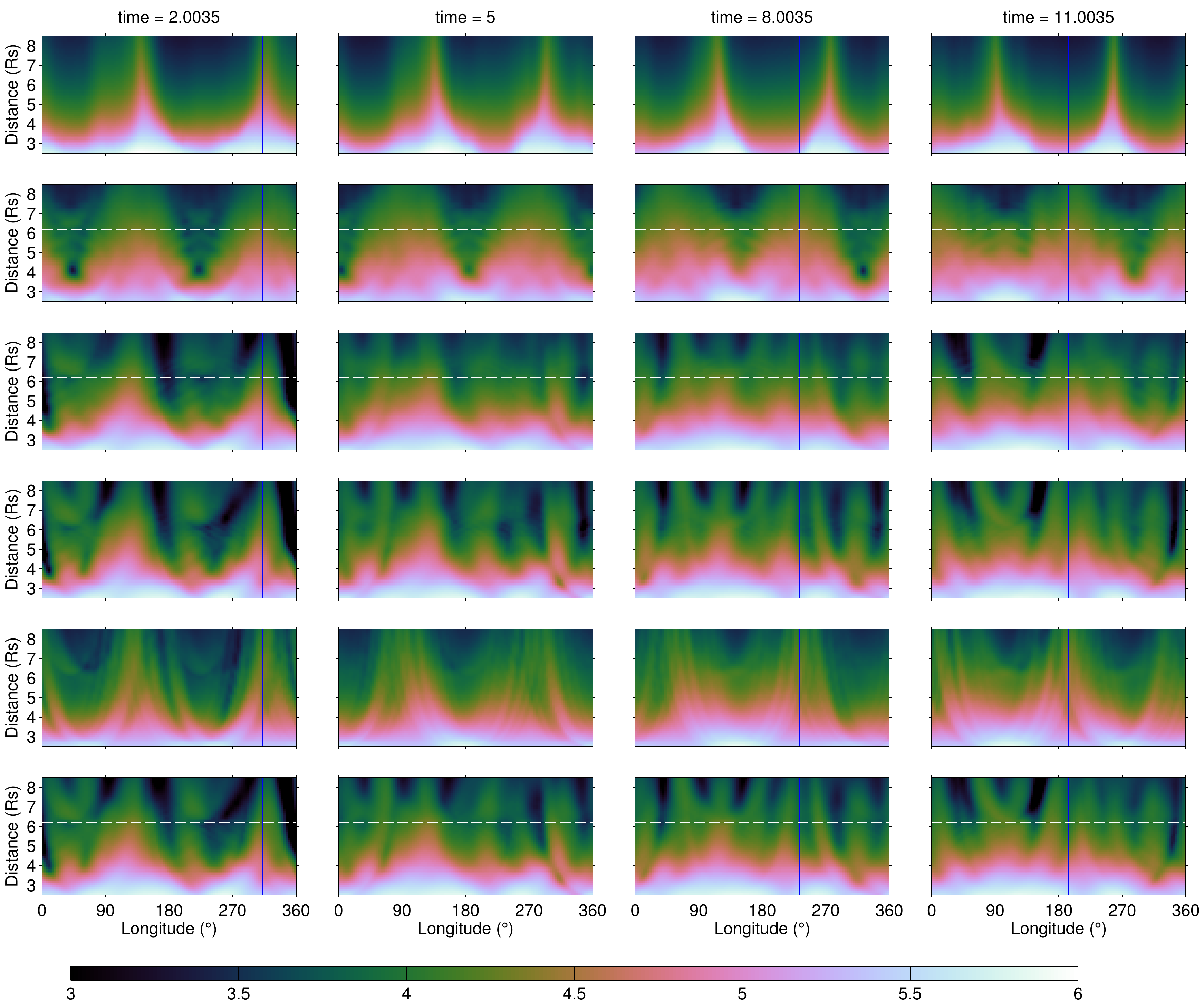}
\caption{Same as Figure~\ref{DynamicSolution_RsRtRc_r405} with the panels representing the equatorial plane in polar coordinates.}
\label{DynamicSolution_RsRtRc_t15}
\end{center}
\end{figure*}

%------------------------------------------------------------------
\begin{tolerant}{1000}
We now test the time-dependent reconstruction with \spatiotemp{} and co-rotating regularizations as expressed by Equation~\eqref{dyn_corot_mle}, with various choices of the temporal and \corotating{} regularization parameters, the spatial regularization being set to its previous optimal value at $\lambda_{s}=\num{2.2E-6}$.
These results are displayed in Figures~\ref{DynamicSolution_RsRtRc_r405} and~\ref{DynamicSolution_RsRtRc_t15}.
\end{tolerant}

We first examine whether the co-rotating regularization may simply replace the temporal regularization, allowing more freedom for the time variation and simultaneously avoiding the concentration in the plane of the sky.
This case is displayed in the second row of both figures.
It appears that this solution is inferior (more ZDA) to the \spatiotemp{} regularization presented in the previous section, see the last row of Figure~\ref{DynamicSolution_RsRt_r405}.
Even with a small amount of temporal regularization (but less than that optimally found for the \spatiotemp{} regularization only), the addition of the co-rotating matrix does not help, see third row.
A temporal regularization at the same level as before (\spatiotemp{} regularization without co-rotating) is thus required. 
As a consequence, the introduction of the co-rotation minimization does not improve the time resolution of the solution. 

% large C => bad, optimal C no big diff than without (on images)
The three last rows in Figures~\ref{DynamicSolution_RsRtRc_r405} and~\ref{DynamicSolution_RsRtRc_t15} display reconstructions with fixed optimal spatial and time regularization parameters ($\lambda_s$\ and $\lambda_t$)  and three very different values of $\lambda_c$: \numlist{1e-10;2e-7;1e-5}.
This experiment shows that a large amount of co-rotating regularization (fifth row) leads to a poor solution, with perhaps less ZDA, but presenting strange features absent in the model and a poorly recovered neutral sheet. 

\begin{figure} % 13
\begin{center}
 \includegraphics[width=\columnwidth]{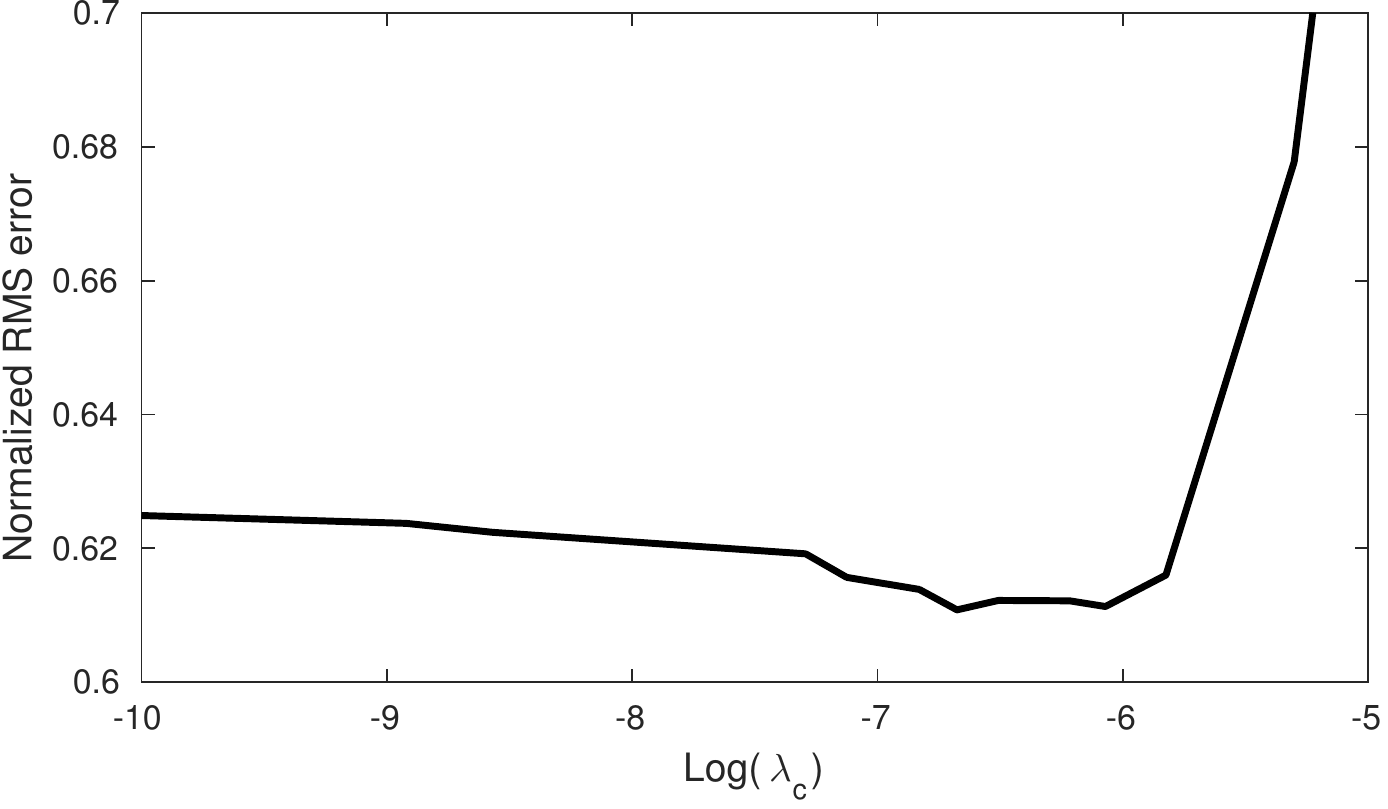}
\caption{RMS error normalized by the standard deviation of the model, as a function of the \corotating{} regularization parameter $\lambda_c$ for fixed values of the spatial and temporal regularization parameters ($\lambda_{s}=\num{2.2e-6}$ and $\lambda_{t}=\num{1.7e-6}$).}
\label{rms_error_lc}
\end{center}
\end{figure}

% better look on RMS error => there is optimal value with small improvement.
To quantitatively assess the effect of this regularization, we plot in Figure~\ref{rms_error_lc} the RMS error normalized by the variance --- see Equation~\eqref{rms_eq} --- averaged over the radial distance to the Sun, as a function of the co-rotating regularization parameter $\lambda_c$ (still with $\lambda_s$ and $\lambda_t$ at their optimal values).
Indeed, a small amount ($\approx \num{2e-7}$) of co-rotating regularization slightly improves the reconstruction.
This case is displayed in the last rows of Figures~\ref{DynamicSolution_RsRtRc_r405} and~\ref{DynamicSolution_RsRtRc_t15} and can be compared with the fourth row with almost no co-rotating regularization.

Figure~\ref{rms_error_lc} shows again that, increasing $\lambda_c$ just above its best value, rapidly degrades the solution to situations even worse than without co-rotating regularization. 
This can be tentatively explained by the fact that too much co-rotating regularization does remove legitimate co-rotating components present in the model, where a co-rotating component is any solution ($x$) that satisfies Equation~(\ref{eq:co-rot}); however, as explained in~\ref{CoRot}, solutions that satisfy azimuthal symmetry are not penalized.  
As the MHD model has lower boundary conditions set by magnetograms, there are likely components of the model that are unwittingly penalized by the regularization matrix we constructed.

\subsubsection{Reconstruction with a Minimal Background Prior}
%-------------------------------------------------------------

We now consider the possible advantage of constraining the solution to a minimal background of the electron density ($N_e$) as described in Section~\ref{sec_min_bkg} using the best parameter values obtained so far, \ie with co-rotating \spatiotemp{} regularization ($\lambda_{s}=\num{2.2e-6}$; $\lambda_{t}=\num{1.7e-6}$ and $\lambda_{c}=\num{2e-7}$).

We consider the axi-symmetric model of \citet{Saito72} for the corona of the minimum type and more precisely, its polar profile to build a spherically symmetric model of the minimal background:

\begin{equation}
 N_e(r) =\biggl[ \frac{1.545}{r^{16}} + \frac{0.079}{r^{6}} \biggr]  \times \SI{E8}{\cmc}\,.
\end{equation}

\begin{figure} % 14
\begin{center}
 \includegraphics[width=\columnwidth]{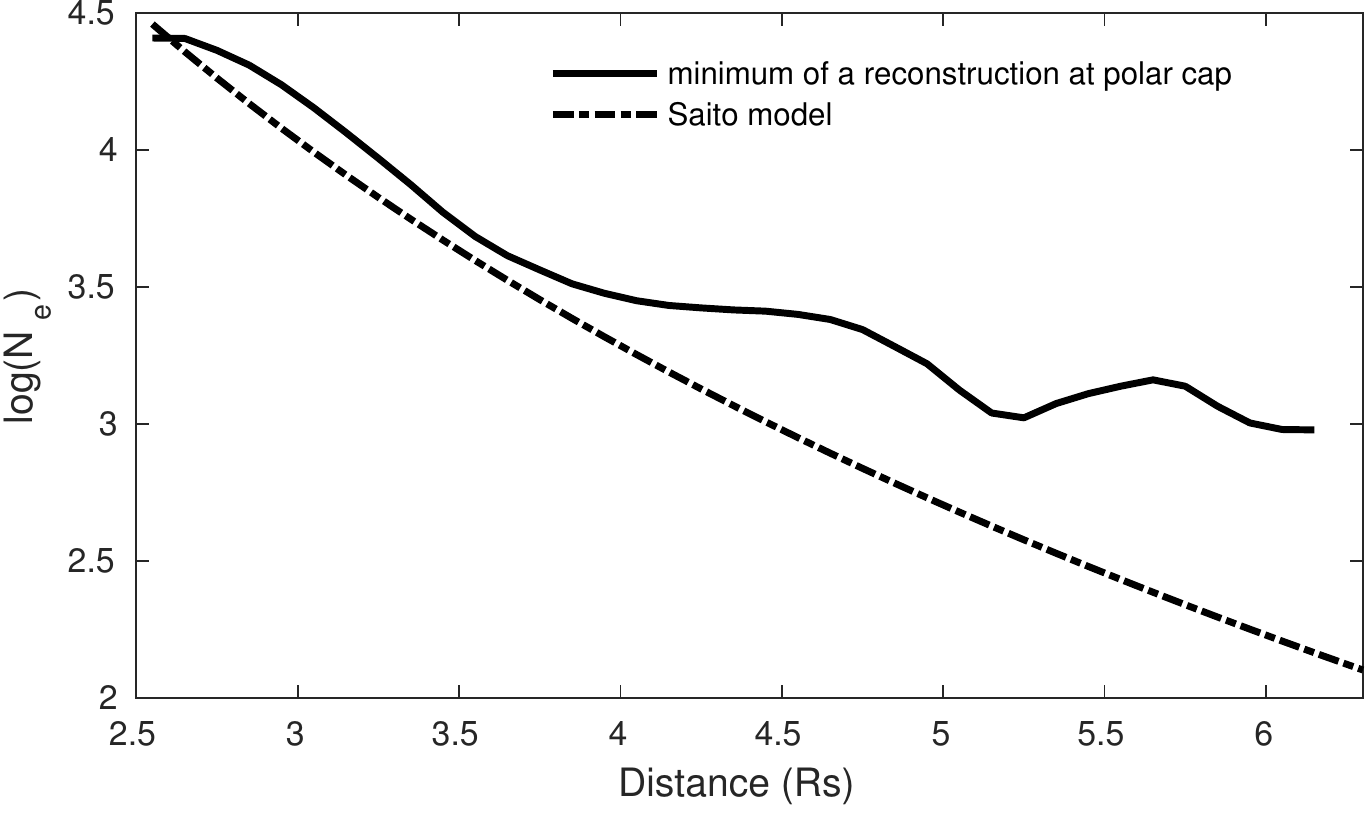}
\caption{Radial profiles of several backgrounds representing the minimum of the coronal radiance. 
The solid line represents the minimum of a reconstruction in radially increasing polar caps and the dotted-dashed line represents polar profile of the axi-symmetric model of \citet{Saito72}.}
\label{back}
\end{center}
\end{figure}

Figure~\ref{back} compares this profile with that resulting from the previous regularized inversion (without the background constraint). This latter profile is obtained by taking the minima of the electron density on successive  polar caps at increasing radius and limited by a small cone centered on the polar axis. 
One can see that the Saito profile does provide a satisfactory minimal background up to \SI{4}{\Rsun}; beyond, the reconstructed profile is affected by noise.

In order to obtain a model that is closer to the data, we can even scale the model in amplitude.  
The scaling coefficient which minimizes the least-square difference between the two is 
\begin{equation}
\hat{a} = \frac{\transp{\bvec{m}}\bvec{p}}{\transp{\bvec{m}}\bvec{m}} \, ,
\end{equation}
where $\bvec{p}$ is the measured minimum $N_e$ radial profile, $\bvec{m}$ the Saito model and the summation extends over the whole radial range.
Using the LASCO-C2 data, we obtain a scaling coefficient of \num{1.14}.
 
\begin{figure*} % 15
\begin{center}
 \includegraphics[width=\textwidth]{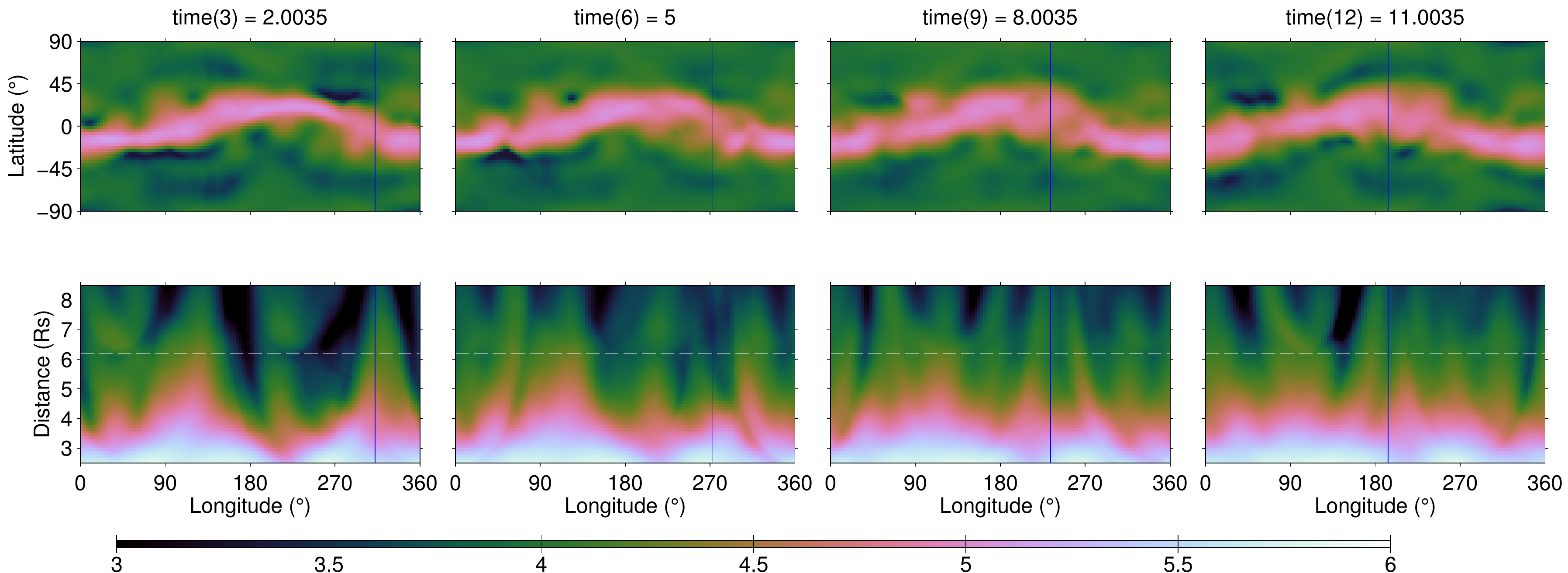}
\caption{A dynamic reconstruction method with a background corona applied to the dynamic model. 
The panels represent slices of the tomographic determinations of the electron density in units of \si{\cmc} using a logarithmic scale.
The first row corresponds to spherical shells with radius of \SI{4.05}{\Rsun} and the second row to the equatorial plane in polar coordinates.}
\label{rebuildSaito}
\end{center}
\end{figure*}

Figure~\ref{rebuildSaito} displays the results of the reconstruction that incorporates the above background. 
They can be compared with the last rows of Figures~\ref{DynamicSolution_RsRtRc_r405} and~\ref{DynamicSolution_RsRtRc_t15}. 
Visually, the images are indistinguishable.   
Indeed, we expect the introduction of a minimal background to essentially replace the zero values in the ZDA to the minimal background values. 
But these patches remain much lower than their surroundings and thus appear as artifacts. 
However, and as it will be addressed later in the discussion, if we compare the RMS errors of reconstructions with and without the background, we do see a slight improvement.

\subsection{Application to the LASCO-C2 Images}
\label{dynrec_lasco}
%---------------------------------------------

\begin{figure*} % 16
(a)\begin{center}
 \includegraphics[width=\textwidth]{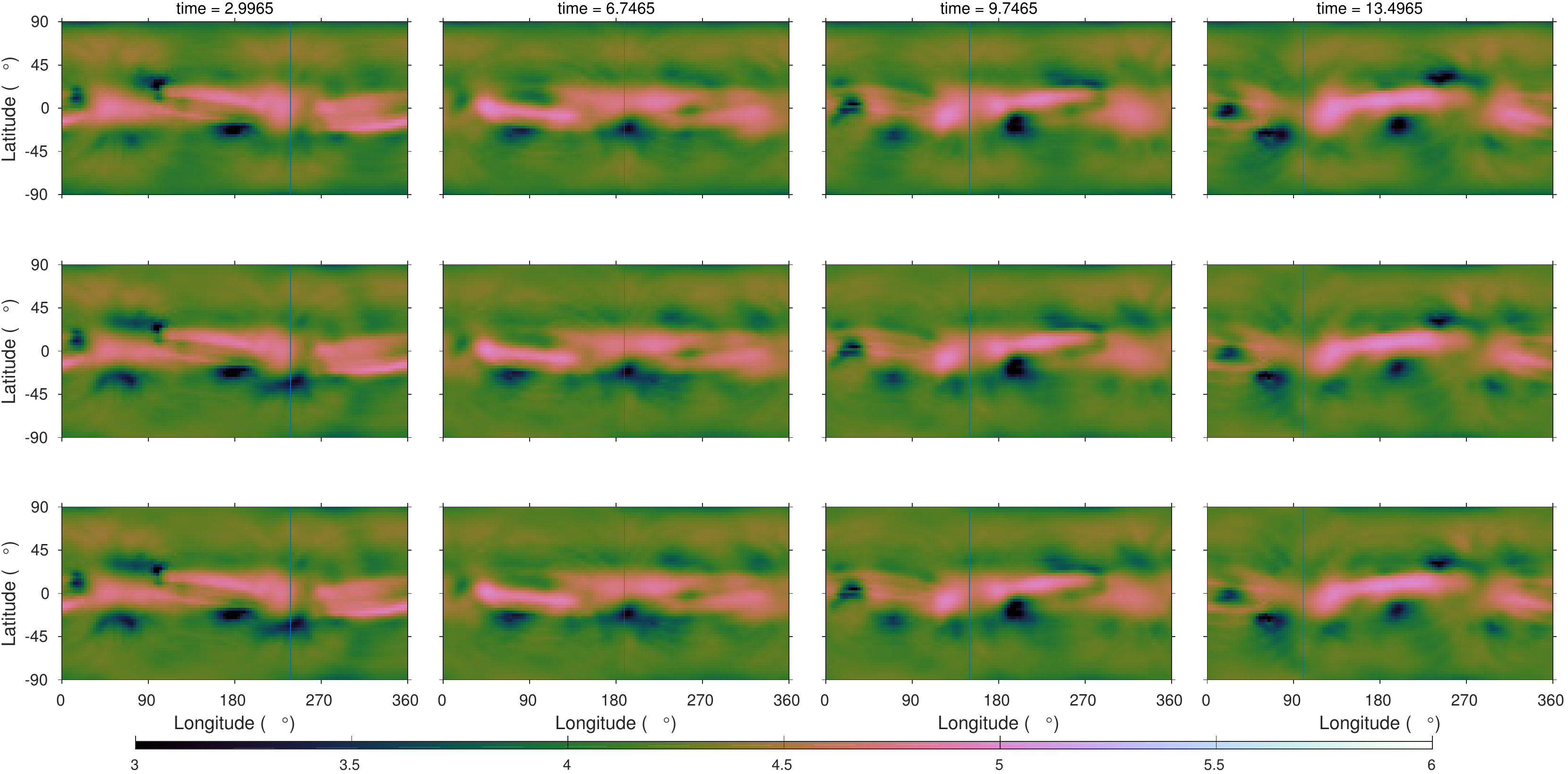}
\end{center}
(b)\begin{center}
 \includegraphics[width=\textwidth]{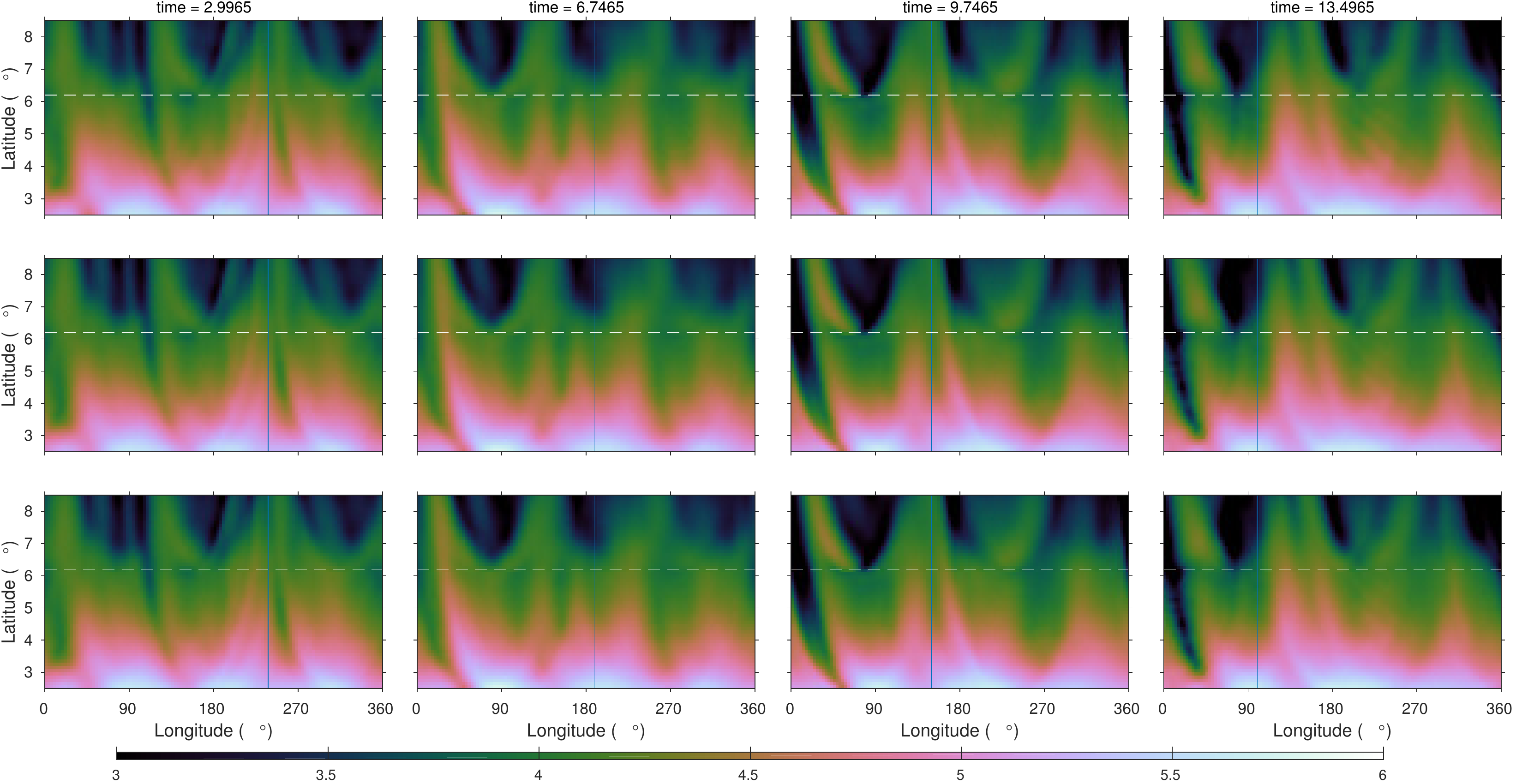}
\end{center}
\caption{(a) Illustration of the \spatiotemp{} regularized dynamic reconstruction of LASCO-C2 images, for the three dynamic reconstruction methods.
The panels represent shells at a constant radius ($r=\SI{4.05}{\Rsun}$) of the electron density in units of \si{\cmc} using a logarithmic scale.
Each column corresponds to a reconstruction of the corona at a given time.
The first (top) row corresponds to a reconstruction with the $\bvec{S}$ and $\bvec{T}$ matrices, the second row to reconstruction with the $\bvec{S}$, $\bvec{T}$ and $\bvec{C}$ matrices and the bottom row to a reconstruction with the $\bvec{S}$, $\bvec{T}$ and $\bvec{C}$ matrices and a background.
The blue vertical line corresponds to the Carrington longitude of the observer.
(b) Same as (a) with the panels representing the equatorial plane in polar coordinates.}
\label{DynLASCO_r}
\end{figure*}

We now  apply the time-dependent tomography to the set of 53 LASCO-C2 $pB$ images described in \ref{lasco} and display the results in Figure~\ref{DynLASCO_r} using the three procedures (\spatiotemp{} regularization + co-rotating regularization + minimal background) optimally tuned on the basis of the MHD model, \ie with the same values for $\lambda_s,\lambda_t,\lambda_c$.

The outcome of the tomographic reconstructions of the LASCO images is similar to that of the MHD model images.
In particular, the reconstruction with the \spatiotemp{} regularization alone (first row) outperforms all static reconstructions (compare with Figure~\ref{SolutionStat_5methods_lasco}). 
Similarly, the addition of a co-rotating regularization is disappointing since it does not improve significantly the reconstruction (second row).
Therefore, the tentative explanation given above in Section~\ref{rec_corot_mhd} of the presence of more intrinsic co-rotating components in the model than in the real corona cannot be invoked to explain the lack of benefit of the co-rotating regularization.

\subsection{Discussion and comparison}
\label{discomp}

\begin{figure*} % 17
\begin{center}
 \includegraphics[width=\textwidth]{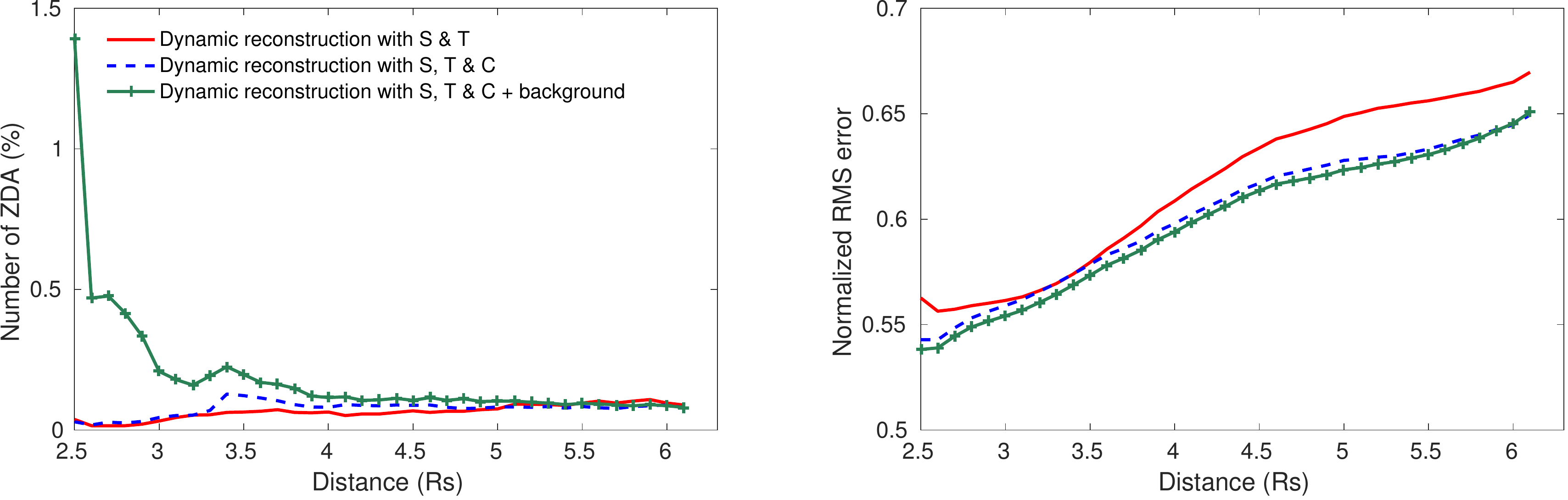}
 \includegraphics[width=\textwidth]{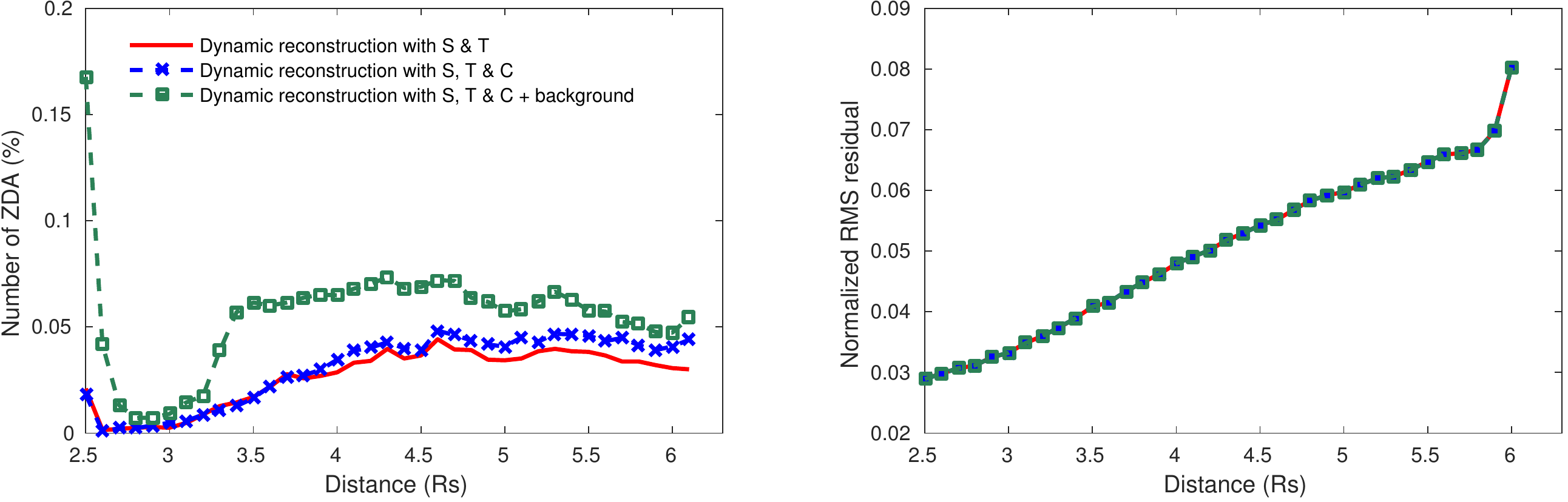}
\caption{Quality of the solution in spherical shells of increasing distance from the Sun for the three dynamic reconstruction methods applied to 14 $pB$ images of the dynamic MHD model (top panels) and to 53 LASCO-C2 $pB$ images (bottom panels).
The two left panels display the percentage of ZDA. 
The top-right panel displays the RMS error normalized by the standard deviation of the model and the bottom-right panel displays the RMS residual error normalized by the standard deviation of the image pixels. In the latter the three curves are superimposed.}
\label{MHDerr}
\end{center}
\end{figure*}

Figure~\ref{MHDerr} presents a quantitative assessment of the reconstructions performed with the MHD model and LASCO images in terms of ZDA number and RMS error normalized by the standard deviation of the model in the former case and RMS residual normalized by the variance of the data in the latter case.
In essence, these figures confirm the qualitative conclusions based on the displayed shells of the time-dependent reconstructions (Figures~\ref{DynamicSolution_RsRt_r405}, \ref{DynamicSolution_RsRtRc_r405}, \ref{DynamicSolution_RsRtRc_t15}, \ref{rebuildSaito}, and \ref{DynLASCO_r}).
The addition of the co-rotating regularization offers only a marginal improvement in both cases, MHD model and LASCO images (same amount of ZDA but very slight decrease of the RMS error). 
The introduction of the background does not provide any noticeable quantitative gain on the reconstruction in both cases as best seen on the RMS error and residual curves. 
The number of ZDA in the two reconstructions is larger with the background but this effect is misleading.
On the one hand, in the absence of background, this number is the percentage of exactly  null voxels which are strictly artifacts since the electron density cannot be exactly zero on physical grounds. 
On the other hand, when the background is included, we only count the voxels at the level of the background and that includes some voxels at physically acceptable density values thus overestimating the number of artifacts.

Comparing now these figures with those corresponding to the static reconstructions (Figures~\ref{MHD_StatZerosRMS} and~\ref{lasco_stat_zeros}), we can appreciate the net improvement resulting from the time-dependent inversion. 
It reduces the number of ZDA by a factor of 30, from $\approx \SI{3}{\percent}$ to $\approx \SI{0.1}{\percent}$ and the RMS error by a factor $\approx 2$.
\citet{Butala10} observed a similar trend when they performed a Kalman time-dependent inversion of STEREO/COR1 data, but the improvement on their number of ZDA was only a factor 2 (from $\approx \SI{5}{\percent}$ to $\approx \SI{2.5}{\percent}$), which is much less than what we have achieved. 
However, we must keep in mind that \citet{Butala10} reconstructed the corona between 1.5 and \SI{3.3}{\Rsun} where one would expect more ZDA due to larger dynamical effects in the innermost corona.

\section{Conclusion}
\label{conclusion}

In this article, we considered the main problem crippling solar rotational tomography, namely the assumption of a stable corona whereas it is intrinsically dynamic even during the minima of solar activity.
Using a dynamic MHD model of the corona, we showed that a static reconstruction of the three-dimensional electron density results in severe artifacts that do not appear when applied to a static corona.
We addressed the problem using two different approaches:
i) mitigation of the temporal variation in the framework of a regularized static inversion, and
ii) a true time-dependent inversion.

Our main results based on tests performed on synthetic images constructed from this dynamic MHD model and on real LASCO-C2 images of the corona  are summarized below.

\begin{itemize}
    \item Crucial to testing our procedure and properly tuning the regularization parameters was the introduction of a time-dependent MHD model of the corona based on observed magnetograms to build a time-series of synthetic images of the corona.

    \item Our mitigation procedures --- multiple masking with simple juxtaposition and recursive combination of solutions, radial weighting with a mean radial intensity profile or an exponential profile --- do not convincingly improve the reconstruction.

    \item A true time-dependent inversion with \spatiotemp{} regularization does improve the situation and convincingly reduces the artifacts by a factor 30 and the normalized RMS error by a factor 2.

    \item The introduction of an additional \spatiotemp{} regularization that penalizes the coronal structures that are apparently fixed for the observer turns out to be disappointing. However it could potentially improve the reconstruction when using more frequent images.

    \item Whereas non-physical densities, although greatly reduced, are still present, our dynamic reconstruction appears qualitatively superior, exhibiting a generally smoother and more connected, \ie more physically reasonable, reconstructed streamer belt.

\end{itemize}

Future work will consider a possible improvement of the co-rotation regularization and will ultimately consist in applying our time-dependent SRT procedure to the whole set of LASCO-C2 images presently extending over 20 years, that is almost two solar cycles.
It is hoped that the resulting four-dimensional estimates of the electron density will provide insight into the coronal and solar wind processes.

%%%%%%%%%%%%%%%%%%%%%%%%%%%%%%%%%%%%%%%%%%%%%%%%%%%%%%%%%%%%%%%%%%%%%%%%%%%

\section*{Acknowledgements}
We thank B.~Boclet and A.~Llebaria for processing the LASCO-C2 $pB$ images.
The LASCO-C2 project at the Laboratoire d'Astro\-phsique de Marseille is funded by the Centre National d'Etudes Spatiales (CNES).
LASCO was built by a consortium of the Naval Research Laboratory, USA, the Laboratoire d'Astrophysique de Marseille (formerly Laboratoire d'Astronomie Spatiale), France, the Max-Planck-Institut f\"ur Sonnensystemforschung (formerly Max Planck Institute f\"ur Aeronomie), Germany, and the School of Physics and Astronomy, University of Birmingham, UK. 
SOHO is a project of international cooperation between ESA and NASA.

%==================================================================================
%%% APPENDIX
%==================================================================================
\appendix

\section{The Time-dependent MHD Model of the Corona}
\label{mhdmodel}
%---------------------------------------------------

In order to simulate a time-dependent solar corona, we used a time-series of steady-state MHD models.  
Each MHD model in the series differs only by the input magnetogram, one per day, for the time period corresponding to CR2077 (November 2008).  
These magnetograms were produced by the data assimilation model of \citet{Upton14}, fed by HMI data.
The MHD model, described by \citet{vanderHolst14}, and known as the Alfv{\'e}n wave solar model (AWSoM), is a global model from the upper chromosphere to the corona and the heliosphere. 
The coronal heating and solar wind acceleration are addressed with low-frequency Alfv{\'e}n wave turbulence. 
The injection of Alfv{\'e}n wave energy at the inner boundary is such that the Poynting flux is proportional to the magnetic field strength. 
The three-dimensional magnetic field topology is simulated using magnetogram data. 
This model does not impose open-closed magnetic field boundaries; those develop self-consistently. 
The physics includes the following: 
\begin{enumerate}
	\item \begin{tolerant}{1000} The Alfv{\'e}n waves are partially reflected by the Alfv{\'e}n speed gradient and the vorticity along the field lines. 
The resulting counter-propagating waves are responsible for the nonlinear turbulent cascade. 
The balanced turbulence due to uncorrelated waves near the apex of the closed field lines and the resulting elevated temperatures are addressed. \end{tolerant}
	\item The turbulent wave dissipation employs the results of the theories of linear wave damping and nonlinear stochastic heating.
	\item The model incorporates collisional and collisionless heat conduction. 
\end{enumerate}

Note that in the time interval between performing the simulations presented here and publication of the article of \citet{vanderHolst14}, the AWSoM was updated to employ three different temperatures, namely the isotropic electron temperature and the parallel and perpendicular ion temperatures.

\begin{figure*} % 21
\begin{center}
 \includegraphics[width=\textwidth]{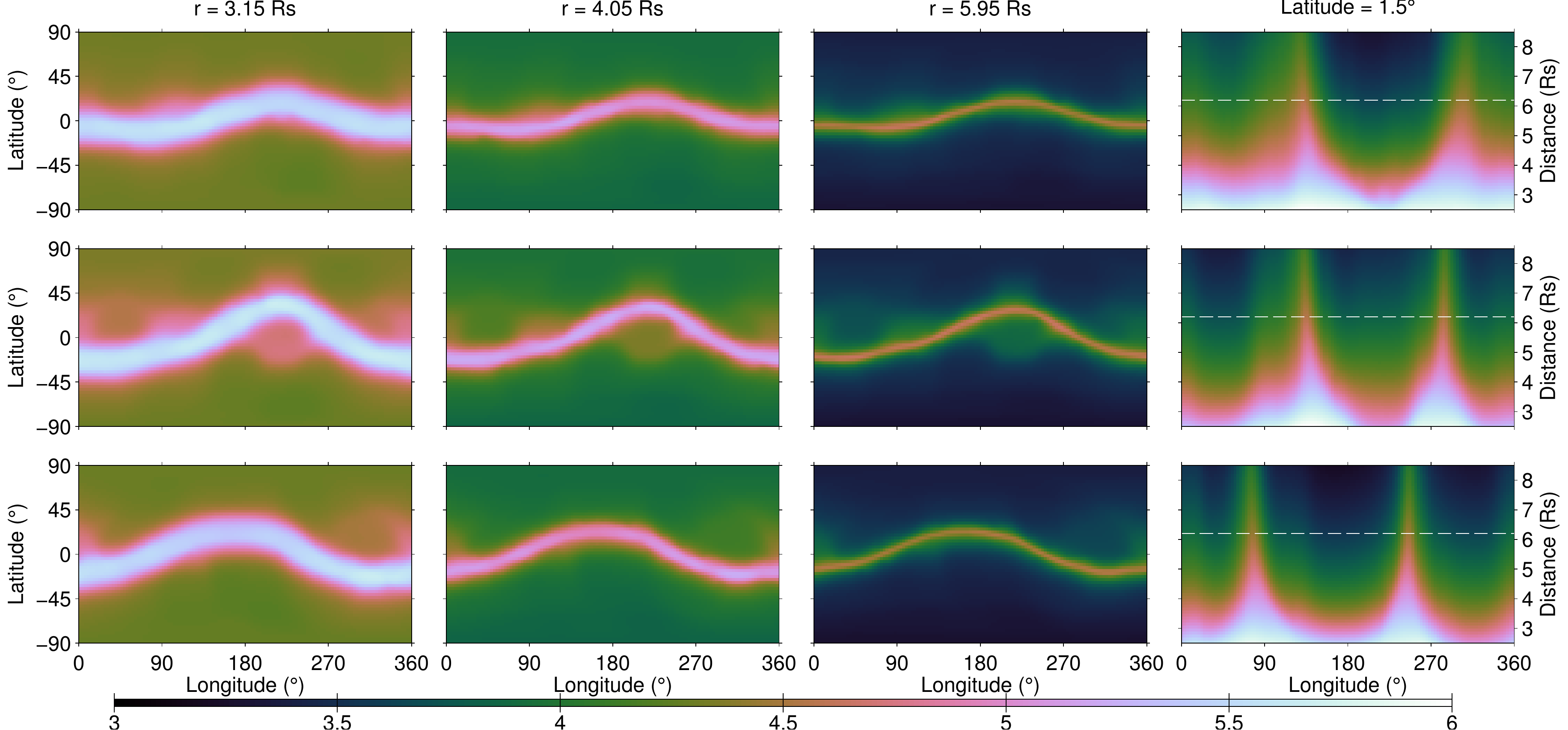}
\caption{Illustration of the dynamic MHD model at different dates: 21/11/2008 (first row), 27/11/2008 (second row), 04/12/2008 (third row).
The 12 panels represent slices of the electron density in units of \si{\cmc} using a logarithmic scale.
The 3 leftmost columns correspond to spherical shells with radii of \SIlist{3.15; 4.05; 5.95}{\Rsun} and the rightmost column to the equatorial plane in polar coordinates.}
\label{modelAnex}
\end{center}
\end{figure*}

The 3D model of the  electron density $N_{e}$ is discretized on a spherical grid which rotates with the Sun and has bins equally spaced in latitude and longitude with a bin size of \SI{1}{\degree} and in radius extending from \SIrange{2.5}{20}{\Rsun} with a bin size of \SI{0.05}{\Rsun}. It is illustrated in Figure~\ref{modelAnex}.

Synthetic images are constructed by sub-sampling the model to a resolution of \SI{3}{\degree} in latitude and longitude and \SI{0.1}{\Rsun} radially, and integrating the Thompson scattering along each \los{} corresponding to each pixel.
They have the same characteristics as the LASCO-C2 images taken in November 2008, same \fov{} (\SIrange{2.5}{6.2}{\Rsun}) and view angles.

\section{The LASCO-C2 Images}
\label{lasco}
%---------------------------

The LASCO-C2 instrument routinely acquires unpolarized images of $1024^2$ pixels and polarization sequences composed of three images with polarizers oriented at \SI{60}{\degree} from each other and an additional unpolarized image of the same format of $512^2$  pixels. 
The cadence of the polarization sequences is typically one per day but has been progressively increased as more telemetry became available.
In addition, high-cadence sequences have been obtained for specific purposes.
Such a sequence at four images per day has been used by \citet{Frazin10} for their tomographic analysis and we retained it for the present tests.
The data set consists of 53 sequences, roughly evenly spaced in time, starting at about 21:00 UT on 15 March 2009 and ending at about 15:00 UT on 29 March, which falls within Carrington Rotation 2081. 
The procedures to correct and calibrate the LASCO-C2 images have however been substantially improved in the meantime.
The most recent description of the pipeline processing of the LASCO-C2 images is given by \citet{Lamy14} and it produces fully corrected and calibrated images of the polarized radiance $pB$ and of the radiance of the K-corona $B_K$.
It incorporates an improved absolute radiometric calibration \citep{Gardes13} and slight corrections of the exposure times. 
Additional corrections have been introduced thereafter, incidentally as a result of finding stray effects in the very first tests of our tomographic reconstruction.
First an error has been detected in the orientation of the images which was corrected and even more important, the complex (because of the presence of the pylon holding the occulter) vignetting function of the coronagraph has been refined. 
As a consequence, the whole set of LASCO-C2 images has been reprocessed.
Whereas LASCO-C2 has a \fov{} extending from \SIrange{2.2}{6.2}{\Rsun}, in practice the inner boundary is set to \SI{2.5}{\Rsun} to avoid interferences from the bright diffraction fringe surrounding the occulter.

Solving the inverse problem to reconstruct the coronal density requires to build the projection matrix ($\mtx{A}$), which is determined by the geometry and the physics of the problem.
For each pixel located between \SI{2.5}{\Rsun} and \SI{6.2}{\Rsun}, the geometry of the \los{} is specified using the information in the headers of the LASCO-C2 images and in the ancillary data, \ie date of observation, attitude and location of the SOHO satellite, heliographic latitude, which is the tilt angle of the ecliptic with respect to the solar equatorial plane also called the $B_{0}$ angle. 
The Thomson scattering function is determined for each \los{} and summed over four adjacent lines-of-sight.
Once we have built the matrix $\mtx{A}$ for the LASCO images of size $512^2$ pixels, we resample the images with a factor 2 to obtain images of size $256^2$ pixels to be used for the reconstruction.
This summation increases the accuracy of the coefficients of the matrix.

\section{Co-rotating Regularization}
\label{CoRot}
%----------------------------------

We first define the operator ($\mtx{R}$) which rotates a static corona $x(r,\theta,\phi)$ around the polar axis by an angle ($\varphi$) via

\begin{equation}
\mtx{R}(\varphi) \bvec{x}(r,\theta,\phi) \equiv \bvec{x}(r,\theta,\phi+\varphi) \,.
\label{eq:co-rot}
\end{equation}

The implementation of such a rotation operator needs some interpolation scheme since the rotation angle ($\varphi$) may be a non-integer longitude shift and we chose a simple linear interpolation.
We then define $\mtx{R}_i$ such that $\mtx{R}_i \equiv \mtx{R}(\varphi_i)$ where $\varphi_i$ is the angle between two successive images taken at times $t_i$ and $t_{i+1}$.
Let $\bvec{x}_i$ be the solution at time $t_i$, then the rotating components satisfy: 
$$\forall i\; \mtx{R}_i \bvec{x}_i  - \bvec{x}_{i+1}=0 $$ 
and are therefore solution of the system $\mtx{D}\bvec{x}=0$, where the linear operator $\mtx{D}$ is:
\begin{equation}
\mtx{D} =
\begin{pmatrix}
\mtx{R}_1 & -\mtx{I}  &          &                 &          \\
          & \mtx{R}_2 & -\mtx{I} &                 &          \\
          &           &  \ddots      & \ddots              &          \\
          &           &          & \mtx{R}_{N_t-1}  &  -\mtx{I} 
\end{pmatrix} \; .
\end{equation}
The regularization operator is then the projection onto the null space of $\mtx{D}$, since it extracts the co-rotating components of a time-dependent solution and will tend to lower them.

This rotation-differential matrix $\mtx{D}$ is redundant and operates only on $(\phi,t)$ slices at any given radius and latitude $(r,\theta)$: $\mtx{D}= \mtx{D}_{\textnormal{slice}} \otimes \mtx{I}_{N_r \times N_\theta}$.
To build a basis of the null space of $\mtx{D}$, we only need to build a basis ($\mtx{L}$) of the null space of $\mtx{D}_{slice}$ ($\ker(\mtx{D}_{\textnormal{slice}})$).
For this, we compute its singular value decomposition: $\mtx{D}_{\textnormal{slice}} = \mtx{U} \mtx{\Sigma} \transp{\mtx{V}}$ and retain the last columns of $\mtx{V}$ corresponding to the null singular values in $\mtx{\Sigma}$. 
We then need to remove from $\ker(\mtx{D}_{\textnormal{slice}})$ the sub-space of $\phi$-invariant vectors because we do not want to penalize them in the reconstruction (there are the constant vectors in the $(\phi,t)$ slices space). 
To do so, we add a constant column vector in front of the basis $\mtx{L}$, orthogonalize it and remove the unchanged first constant column vector.
Retaining the same notation ($\mtx{L}$) for this new basis, the desired regularization operator ($\mtx{C}$) is the projection on the sub-space spanned by $\mtx{L}$:

\begin{equation}
 \mtx{C} = \mtx{L} \transp{\mtx{L}} \otimes \mtx{I}_{N_r \times N_\theta} \,.
\end{equation}

In the latter formulation, the co-rotating penalization is computed for each couple of latitude and radius $(r,\theta)$ and the L2 norm of the total is minimized. 
We find preferable to first sum the candidate solution ($\bvec{x}$) along $\theta$ and then apply the co-rotating penalization to a $(\phi,t)$ slice summed over $\theta$ for each radius $r$ :
\begin{equation}
 \mtx{C}_{\Sigma\theta} = [\mtx{L} \transp{\mtx{L}} (\bvec{I}_{N_\phi \times N_t} \otimes \transp{\bvec{1}_{N_\theta}})] \otimes \mtx{I}_{N_r} \,.
\end{equation}

In order to further improve this regularization, it is possible to reduce the null-space of $\mtx{D}$ which is quite large. 
In particular, the above operator penalizes all components that synchronously rotates with the observer and we could try to select a more specific fraction of them.
We experimented in this direction using Fourier filtering along the $\phi$ coordinate and extracting only the low longitudinal frequency modes which rotate synchronously, for instance taking only the even modes (or a fraction of them) corresponding to the components having a \SI{180}{\degree} longitudinal periodicity plus some of its harmonics.
This did not make much difference when applied to the MHD model.   
Instead of operating in the Fourier space, it may be simpler trying to penalize the components that shift from being concentrated in plane $i$ to plane $i+1$ for all $i$, with a weighting function decreasing with the angular distance to this plane.

%==================================================================================
%%% BIBLIOGRAPHY %%%%%%%%%%%%%%%%%%%%%%%%%%%%%%%%%%%%%%%%%%%%%%%%%%%%%%%%%%%
%=================================================================================

\section*{References}
\bibliographystyle{elsarticle-harv}
\bibliography{articles}

\begin{thebibliography}{40}
\expandafter\ifx\csname natexlab\endcsname\relax\def\natexlab#1{#1}\fi
\providecommand{\url}[1]{\texttt{#1}}
\providecommand{\href}[2]{#2}
\providecommand{\path}[1]{#1}
\providecommand{\DOIprefix}{doi:}
\providecommand{\ArXivprefix}{arXiv:}
\providecommand{\URLprefix}{URL: }
\providecommand{\Pubmedprefix}{pmid:}
\providecommand{\doi}[1]{\href{http://dx.doi.org/#1}{\path{#1}}}
\providecommand{\Pubmed}[1]{\href{pmid:#1}{\path{#1}}}
\providecommand{\bibinfo}[2]{#2}
\ifx\xfnm\relax \def\xfnm[#1]{\unskip,\space#1}\fi
%Type = Article
\bibitem[{Allen(1974)}]{Allen74}
\bibinfo{author}{Allen, D.M.}, \bibinfo{year}{1974}.
\newblock \bibinfo{title}{{The Relationship Between Variable Selection and Data
  Agumentation and a Method for Prediction}}.
\newblock \bibinfo{journal}{Technometrics} \bibinfo{volume}{16},
  \bibinfo{pages}{125--127}.
\newblock \DOIprefix\doi{10.1080/00401706.1974.10489157}.
%Type = Article
\bibitem[{{Barbey} et~al.(2008){Barbey}, {Auch{\`e}re}, {Rodet} and
  {Vial}}]{Barbey08}
\bibinfo{author}{{Barbey}, N.}, \bibinfo{author}{{Auch{\`e}re}, F.},
  \bibinfo{author}{{Rodet}, T.}, \bibinfo{author}{{Vial}, J.C.},
  \bibinfo{year}{2008}.
\newblock \bibinfo{title}{{A Time-Evolving 3D Method Dedicated to the
  Reconstruction of Solar Plumes and Results Using Extreme Ultraviolet Data}}.
\newblock \bibinfo{journal}{\solphys} \bibinfo{volume}{248},
  \bibinfo{pages}{409--423}.
\newblock \DOIprefix\doi{10.1007/s11207-008-9151-6},
  \href{http://arxiv.org/abs/0802.0113}{{\tt arXiv:0802.0113}}.
%Type = Article
\bibitem[{{Barbey} et~al.(2013){Barbey}, {Guennou} and
  {Auch{\`e}re}}]{Barbey13}
\bibinfo{author}{{Barbey}, N.}, \bibinfo{author}{{Guennou}, C.},
  \bibinfo{author}{{Auch{\`e}re}, F.}, \bibinfo{year}{2013}.
\newblock \bibinfo{title}{{TomograPy: A Fast, Instrument-Independent, Solar
  Tomography Software}}.
\newblock \bibinfo{journal}{\solphys} \bibinfo{volume}{283},
  \bibinfo{pages}{227--245}.
\newblock \DOIprefix\doi{10.1007/s11207-011-9792-8}.
%Type = Article
\bibitem[{Bro and De~Jong(1997)}]{Bro97}
\bibinfo{author}{Bro, R.}, \bibinfo{author}{De~Jong, S.}, \bibinfo{year}{1997}.
\newblock \bibinfo{title}{{A fast non-negativity-constrained least squares
  algorithm}}.
\newblock \bibinfo{journal}{Journal of chemometrics} \bibinfo{volume}{11},
  \bibinfo{pages}{393--401}.
\newblock
  \DOIprefix\doi{10.1002/(SICI)1099-128X(199709/10)11:5<393::AID-CEM483>3.0.CO;2-L}.
%Type = Article
\bibitem[{{Butala} et~al.(2010){Butala}, {Hewett}, {Frazin} and
  {Kamalabadi}}]{Butala10}
\bibinfo{author}{{Butala}, M.D.}, \bibinfo{author}{{Hewett}, R.J.},
  \bibinfo{author}{{Frazin}, R.A.}, \bibinfo{author}{{Kamalabadi}, F.},
  \bibinfo{year}{2010}.
\newblock \bibinfo{title}{{Dynamic Three-Dimensional Tomography of the Solar
  Corona}}.
\newblock \bibinfo{journal}{\solphys} \bibinfo{volume}{262},
  \bibinfo{pages}{495--509}.
\newblock \DOIprefix\doi{10.1007/s11207-010-9536-1}.
%Type = Article
\bibitem[{Byrd et~al.(1995)Byrd, Lu, Nocedal and Zhu}]{Byrd95}
\bibinfo{author}{Byrd, R.H.}, \bibinfo{author}{Lu, P.},
  \bibinfo{author}{Nocedal, J.}, \bibinfo{author}{Zhu, C.},
  \bibinfo{year}{1995}.
\newblock \bibinfo{title}{{A limited memory algorithm for bound constrained
  optimization}}.
\newblock \bibinfo{journal}{SIAM Journal on Scientific Computing}
  \bibinfo{volume}{16}, \bibinfo{pages}{1190--1208}.
\newblock \DOIprefix\doi{10.1137/0916069}.
%Type = Article
\bibitem[{Floyd et~al.(2014)Floyd, Lamy and Llebaria}]{Floyd14}
\bibinfo{author}{Floyd, O.}, \bibinfo{author}{Lamy, P.},
  \bibinfo{author}{Llebaria, A.}, \bibinfo{year}{2014}.
\newblock \bibinfo{title}{{The Interaction Between Coronal Mass Ejections and
  Streamers: A Statistical View over 15 Years (1996 – 2010)}}.
\newblock \bibinfo{journal}{\solphys} \bibinfo{volume}{289},
  \bibinfo{pages}{1313--1339}.
\newblock \DOIprefix\doi{10.1007/s11207-013-0379-4}.
%Type = Article
\bibitem[{{Frazin}(2000)}]{Frazin00}
\bibinfo{author}{{Frazin}, R.A.}, \bibinfo{year}{2000}.
\newblock \bibinfo{title}{{Tomography of the Solar Corona. I. A Robust,
  Regularized, Positive Estimation Method}}.
\newblock \bibinfo{journal}{\apj} \bibinfo{volume}{530},
  \bibinfo{pages}{1026--1035}.
\newblock \DOIprefix\doi{10.1086/308412}.
%Type = Article
\bibitem[{Frazin et~al.(2005)Frazin, Butala, Kemball and
  Kamalabadi}]{Frazin05b}
\bibinfo{author}{Frazin, R.A.}, \bibinfo{author}{Butala, M.D.},
  \bibinfo{author}{Kemball, A.}, \bibinfo{author}{Kamalabadi, F.},
  \bibinfo{year}{2005}.
\newblock \bibinfo{title}{Time-dependent reconstruction of nonstationary
  objects with tomographic or interferometric measurements}.
\newblock \bibinfo{journal}{The Astrophysical Journal Letters}
  \bibinfo{volume}{635}, \bibinfo{pages}{L197}.
\newblock \DOIprefix\doi{10.1086/499431}.
%Type = Article
\bibitem[{{Frazin} and {Janzen}(2002)}]{Frazin02}
\bibinfo{author}{{Frazin}, R.A.}, \bibinfo{author}{{Janzen}, P.},
  \bibinfo{year}{2002}.
\newblock \bibinfo{title}{{Tomography of the Solar Corona. II. Robust,
  Regularized, Positive Estimation of the Three-dimensional Electron Density
  Distribution from LASCO-C2 Polarized White-Light Images}}.
\newblock \bibinfo{journal}{\apj} \bibinfo{volume}{570},
  \bibinfo{pages}{408--422}.
\newblock \DOIprefix\doi{10.1086/339572}.
%Type = Article
\bibitem[{{Frazin} and {Kamalabadi}(2005)}]{Frazin05a}
\bibinfo{author}{{Frazin}, R.A.}, \bibinfo{author}{{Kamalabadi}, F.},
  \bibinfo{year}{2005}.
\newblock \bibinfo{title}{{Rotational Tomography For 3d Reconstruction Of The
  White-Light And Euv Corona In The Post-Soho Era}}.
\newblock \bibinfo{journal}{\solphys} \bibinfo{volume}{228},
  \bibinfo{pages}{219--237}.
\newblock \DOIprefix\doi{10.1007/s11207-005-2764-0}.
%Type = Article
\bibitem[{{Frazin} et~al.(2010){Frazin}, {Lamy}, {Llebaria} and
  {V{\'a}squez}}]{Frazin10}
\bibinfo{author}{{Frazin}, R.A.}, \bibinfo{author}{{Lamy}, P.},
  \bibinfo{author}{{Llebaria}, A.}, \bibinfo{author}{{V{\'a}squez}, A.M.},
  \bibinfo{year}{2010}.
\newblock \bibinfo{title}{{Three-Dimensional Electron Density from Tomographic
  Analysis of LASCO-C2 Images of the K-Corona Total Brightness}}.
\newblock \bibinfo{journal}{\solphys} \bibinfo{volume}{265},
  \bibinfo{pages}{19--30}.
\newblock \DOIprefix\doi{10.1007/s11207-010-9557-9}.
%Type = Article
\bibitem[{{Frazin} et~al.(2009){Frazin}, {V{\'a}squez} and
  {Kamalabadi}}]{Frazin09}
\bibinfo{author}{{Frazin}, R.A.}, \bibinfo{author}{{V{\'a}squez}, A.M.},
  \bibinfo{author}{{Kamalabadi}, F.}, \bibinfo{year}{2009}.
\newblock \bibinfo{title}{{Quantitative, Three-dimensional Analysis of the
  Global Corona with Multi-spacecraft Differential Emission Measure
  Tomography}}.
\newblock \bibinfo{journal}{\apj} \bibinfo{volume}{701},
  \bibinfo{pages}{547--560}.
\newblock \DOIprefix\doi{10.1088/0004-637X/701/1/547}.
%Type = Article
\bibitem[{{Frazin} et~al.(2007){Frazin}, {V{\'a}squez}, {Kamalabadi} and
  {Park}}]{Frazin07}
\bibinfo{author}{{Frazin}, R.A.}, \bibinfo{author}{{V{\'a}squez}, A.M.},
  \bibinfo{author}{{Kamalabadi}, F.}, \bibinfo{author}{{Park}, H.},
  \bibinfo{year}{2007}.
\newblock \bibinfo{title}{{Three-dimensional Tomographic Analysis of a
  High-Cadence LASCO-C2 Polarized Brightness Sequence}}.
\newblock \bibinfo{journal}{\apjl} \bibinfo{volume}{671},
  \bibinfo{pages}{L201--L204}.
\newblock \DOIprefix\doi{10.1086/525017}.
%Type = Article
\bibitem[{{Frazin} et~al.(2012){Frazin}, {V{\'a}squez}, {Thompson}, {Hewett},
  {Lamy}, {Llebaria}, {Vourlidas} and {Burkepile}}]{Frazin12}
\bibinfo{author}{{Frazin}, R.A.}, \bibinfo{author}{{V{\'a}squez}, A.M.},
  \bibinfo{author}{{Thompson}, W.T.}, \bibinfo{author}{{Hewett}, R.J.},
  \bibinfo{author}{{Lamy}, P.}, \bibinfo{author}{{Llebaria}, A.},
  \bibinfo{author}{{Vourlidas}, A.}, \bibinfo{author}{{Burkepile}, J.},
  \bibinfo{year}{2012}.
\newblock \bibinfo{title}{{Intercomparison of the LASCO-C2, SECCHI-COR1,
  SECCHI-COR2, and Mk4 Coronagraphs}}.
\newblock \bibinfo{journal}{\solphys} \bibinfo{volume}{280},
  \bibinfo{pages}{273--293}.
\newblock \DOIprefix\doi{10.1007/s11207-012-0028-3}.
%Type = Article
\bibitem[{Gard\`{e}s et~al.(2013)Gard\`{e}s, Lamy and Llebaria}]{Gardes13}
\bibinfo{author}{Gard\`{e}s, B.}, \bibinfo{author}{Lamy, P.},
  \bibinfo{author}{Llebaria, A.}, \bibinfo{year}{2013}.
\newblock \bibinfo{title}{{Photometric Calibration of the LASCO-C2 Coronagraph
  over 14 Years (1996 – 2009)}}.
\newblock \bibinfo{journal}{\solphys} \bibinfo{volume}{283},
  \bibinfo{pages}{667--690}.
\newblock \DOIprefix\doi{10.1007/s11207-013-0240-9}.
%Type = Article
\bibitem[{Golub et~al.(1979)Golub, Heath and Wahba}]{Golub79}
\bibinfo{author}{Golub, G.H.}, \bibinfo{author}{Heath, M.},
  \bibinfo{author}{Wahba, G.}, \bibinfo{year}{1979}.
\newblock \bibinfo{title}{{Generalized cross-validation as a method for
  choosing a good ridge parameter}}.
\newblock \bibinfo{journal}{Technometrics} \bibinfo{volume}{21},
  \bibinfo{pages}{215--223}.
\newblock \DOIprefix\doi{10.2307/1268518}.
%Type = Inproceedings
\bibitem[{Golub and Von~Matt(1997)}]{Golub97a}
\bibinfo{author}{Golub, G.H.}, \bibinfo{author}{Von~Matt, U.},
  \bibinfo{year}{1997}.
\newblock \bibinfo{title}{Tikhonov regularization for large scale problems},
  in: \bibinfo{booktitle}{Workshop on Scientific Computing}, pp.
  \bibinfo{pages}{3--26}.
%Type = Article
\bibitem[{Hansen(1992)}]{Hansen92}
\bibinfo{author}{Hansen, P.C.}, \bibinfo{year}{1992}.
\newblock \bibinfo{title}{{Analysis of Discrete Ill-Posed Problems by Means of
  the L-Curve}}.
\newblock \bibinfo{journal}{SIAM Review} \bibinfo{volume}{34},
  \bibinfo{pages}{561--580}.
\newblock \DOIprefix\doi{10.1137/1034115}.
%Type = Article
\bibitem[{Hansen and O’Leary(1993)}]{Hansen93}
\bibinfo{author}{Hansen, P.C.}, \bibinfo{author}{O’Leary, D.P.},
  \bibinfo{year}{1993}.
\newblock \bibinfo{title}{{The Use of the L-Curve in the Regularization of
  Discrete Ill-Posed Problems}}.
\newblock \bibinfo{journal}{SIAM Journal on Scientific Computing}
  \bibinfo{volume}{14}, \bibinfo{pages}{1487--1503}.
\newblock \DOIprefix\doi{10.1137/0914086}.
%Type = Article
\bibitem[{van~der Holst et~al.(2014)van~der Holst, Sokolov, Meng, Jin,
  {Manchester, IV}, T\'{o}th and Gombosi}]{vanderHolst14}
\bibinfo{author}{van~der Holst, B.}, \bibinfo{author}{Sokolov, I.V.},
  \bibinfo{author}{Meng, X.}, \bibinfo{author}{Jin, M.},
  \bibinfo{author}{{Manchester, IV}, W.B.}, \bibinfo{author}{T\'{o}th, G.},
  \bibinfo{author}{Gombosi, T.I.}, \bibinfo{year}{2014}.
\newblock \bibinfo{title}{Alfv\'{e}n wave solar model (awsom): coronal
  heating}.
\newblock \bibinfo{journal}{The Astrophysical Journal} \bibinfo{volume}{782},
  \bibinfo{pages}{81}.
\newblock \DOIprefix\doi{10.1088/0004-637X/782/2/81}.
%Type = Article
\bibitem[{{Huang} et~al.(2012){Huang}, {Frazin}, {Landi}, {Manchester},
  {V{\'a}squez} and {Gombosi}}]{Huang12}
\bibinfo{author}{{Huang}, Z.}, \bibinfo{author}{{Frazin}, R.A.},
  \bibinfo{author}{{Landi}, E.}, \bibinfo{author}{{Manchester}, W.B.},
  \bibinfo{author}{{V{\'a}squez}, A.M.}, \bibinfo{author}{{Gombosi}, T.I.},
  \bibinfo{year}{2012}.
\newblock \bibinfo{title}{{Newly Discovered Global Temperature Structures in
  the Quiet Sun at Solar Minimum}}.
\newblock \bibinfo{journal}{\apj} \bibinfo{volume}{755}, \bibinfo{pages}{86}.
\newblock \DOIprefix\doi{10.1088/0004-637X/755/2/86},
  \href{http://arxiv.org/abs/1207.6661}{{\tt arXiv:1207.6661}}.
%Type = Article
\bibitem[{Kramar et~al.(2014)Kramar, Airapetian, Miki\'{c} and
  Davila}]{Kramar14}
\bibinfo{author}{Kramar, M.}, \bibinfo{author}{Airapetian, V.},
  \bibinfo{author}{Miki\'{c}, Z.}, \bibinfo{author}{Davila, J.},
  \bibinfo{year}{2014}.
\newblock \bibinfo{title}{{3D Coronal Density Reconstruction and Retrieving the
  Magnetic Field Structure during Solar Minimum}}.
\newblock \bibinfo{journal}{Solar Physics} \bibinfo{volume}{289},
  \bibinfo{pages}{2927--2944}.
\newblock \DOIprefix\doi{10.1007/s11207-014-0525-7},
  \href{http://arxiv.org/abs/1405.0951}{{\tt arXiv:1405.0951}}.
%Type = Article
\bibitem[{{Kramar} et~al.(2009){Kramar}, {Jones}, {Davila}, {Inhester} and
  {Mierla}}]{Kramar09}
\bibinfo{author}{{Kramar}, M.}, \bibinfo{author}{{Jones}, S.},
  \bibinfo{author}{{Davila}, J.}, \bibinfo{author}{{Inhester}, B.},
  \bibinfo{author}{{Mierla}, M.}, \bibinfo{year}{2009}.
\newblock \bibinfo{title}{{On the Tomographic Reconstruction of the 3D Electron
  Density for the Solar Corona from STEREO COR1 Data}}.
\newblock \bibinfo{journal}{\solphys} \bibinfo{volume}{259},
  \bibinfo{pages}{109--121}.
\newblock \DOIprefix\doi{10.1007/s11207-009-9401-2}.
%Type = Article
\bibitem[{Lamy et~al.(2014)Lamy, Barlyaeva, Llebaria and Floyd}]{Lamy14}
\bibinfo{author}{Lamy, P.}, \bibinfo{author}{Barlyaeva, T.},
  \bibinfo{author}{Llebaria, A.}, \bibinfo{author}{Floyd, O.},
  \bibinfo{year}{2014}.
\newblock \bibinfo{title}{{Comparing the solar minima of cycles 22/23 and
  23/24: The view from LASCO white light coronal images}}.
\newblock \bibinfo{journal}{Journal of Geophysical Research: Space Physics}
  \bibinfo{volume}{119}, \bibinfo{pages}{47--58}.
\newblock \DOIprefix\doi{10.1002/2013JA019468}.
%Type = Inbook
\bibitem[{Lawson and Hanson(1995)}]{Lawson95}
\bibinfo{author}{Lawson, C.L.}, \bibinfo{author}{Hanson, R.J.},
  \bibinfo{year}{1995}.
\newblock \bibinfo{title}{{23. Linear Least Squares with Linear Inequality
  Constraints}}. chapter~\bibinfo{chapter}{23}.
\newblock pp. \bibinfo{pages}{158--173}.
\newblock \DOIprefix\doi{10.1137/1.9781611971217.ch23}.
%Type = Article
\bibitem[{Louis and Natterer(1983)}]{Louis83}
\bibinfo{author}{Louis, A.K.}, \bibinfo{author}{Natterer, F.},
  \bibinfo{year}{1983}.
\newblock \bibinfo{title}{{Mathematical problems of computerized tomography}}.
\newblock \bibinfo{journal}{Proceedings of the IEEE} \bibinfo{volume}{71},
  \bibinfo{pages}{379--389}.
\newblock \DOIprefix\doi{10.1109/PROC.1983.12596}.
%Type = Article
\bibitem[{Morales and Nocedal(2011)}]{Morales11}
\bibinfo{author}{Morales, J.L.}, \bibinfo{author}{Nocedal, J.},
  \bibinfo{year}{2011}.
\newblock \bibinfo{title}{{Remark on “algorithm 778: L-BFGS-B: Fortran
  subroutines for large-scale bound constrained optimization”}}.
\newblock \bibinfo{journal}{ACM Transactions on Mathematical Software}
  \bibinfo{volume}{38}, \bibinfo{pages}{1--4}.
\newblock \DOIprefix\doi{10.1145/2049662.2049669}.
%Type = Article
\bibitem[{{Nuevo} et~al.(2013){Nuevo}, {Huang}, {Frazin}, {Manchester}, {Jin}
  and {V{\'a}squez}}]{Nuevo13}
\bibinfo{author}{{Nuevo}, F.A.}, \bibinfo{author}{{Huang}, Z.},
  \bibinfo{author}{{Frazin}, R.}, \bibinfo{author}{{Manchester}, iv, W.B.},
  \bibinfo{author}{{Jin}, M.}, \bibinfo{author}{{V{\'a}squez}, A.M.},
  \bibinfo{year}{2013}.
\newblock \bibinfo{title}{{Evolution of the Global Temperature Structure of the
  Solar Corona during the Minimum between Solar Cycles 23 and 24}}.
\newblock \bibinfo{journal}{\apj} \bibinfo{volume}{773}, \bibinfo{pages}{9}.
\newblock \DOIprefix\doi{10.1088/0004-637X/773/1/9}.
%Type = Article
\bibitem[{Orieux et~al.(2013)Orieux, Giovannelli, Rodet and Abergel}]{Orieux13}
\bibinfo{author}{Orieux, F.}, \bibinfo{author}{Giovannelli, J.F.},
  \bibinfo{author}{Rodet, T.}, \bibinfo{author}{Abergel, A.},
  \bibinfo{year}{2013}.
\newblock \bibinfo{title}{{Estimating hyperparameters and instrument parameters
  in regularized inversion Illustration for Herschel /SPIRE map making}}.
\newblock \bibinfo{journal}{Astronomy \& Astrophysics} \bibinfo{volume}{549},
  \bibinfo{pages}{A83}.
\newblock \DOIprefix\doi{10.1051/0004-6361/201219950}.
%Type = Article
\bibitem[{Paige and Saunders(1982)}]{Paige82}
\bibinfo{author}{Paige, C.C.}, \bibinfo{author}{Saunders, M.a.},
  \bibinfo{year}{1982}.
\newblock \bibinfo{title}{{LSQR: An Algorithm for Sparse Linear Equations and
  Sparse Least Squares}}.
\newblock \bibinfo{journal}{ACM Transactions on Mathematical Software}
  \bibinfo{volume}{8}, \bibinfo{pages}{43--71}.
\newblock \DOIprefix\doi{10.1145/355984.355989}.
%Type = Article
\bibitem[{{Saito} et~al.(1970){Saito}, {Makita}, {Nishi} and {Hata}}]{Saito72}
\bibinfo{author}{{Saito}, K.}, \bibinfo{author}{{Makita}, M.},
  \bibinfo{author}{{Nishi}, K.}, \bibinfo{author}{{Hata}, S.},
  \bibinfo{year}{1970}.
\newblock \bibinfo{title}{{A non-spherical axisymmetric model of the solar K
  corona of the minimum type.}}
\newblock \bibinfo{journal}{Annals of the Tokyo Astronomical Observatory}
  \bibinfo{volume}{12}, \bibinfo{pages}{53--120}.
%Type = Article
\bibitem[{Stark and Parker(1995)}]{Stark95}
\bibinfo{author}{Stark, P.B.}, \bibinfo{author}{Parker, R.L.},
  \bibinfo{year}{1995}.
\newblock \bibinfo{title}{Bounded-variable least-squares: an algorithm and
  applications}.
\newblock \bibinfo{journal}{Computational Statistics} \bibinfo{volume}{10},
  \bibinfo{pages}{129--129}.
%Type = Article
\bibitem[{Stone(1974)}]{Stone74}
\bibinfo{author}{Stone, M.}, \bibinfo{year}{1974}.
\newblock \bibinfo{title}{{Cross-Validatory Choice and Assessment of
  Statistical Predictions}}.
\newblock \bibinfo{journal}{Journal of the Royal Statistical Society. Series B
  (Methodological)} \bibinfo{volume}{36}, \bibinfo{pages}{111--147}.
%Type = Article
\bibitem[{Upton and Hathaway(2014)}]{Upton14}
\bibinfo{author}{Upton, L.}, \bibinfo{author}{Hathaway, D.H.},
  \bibinfo{year}{2014}.
\newblock \bibinfo{title}{Predicting the sun's polar magnetic fields with a
  surface flux transport model}.
\newblock \bibinfo{journal}{The Astrophysical Journal} \bibinfo{volume}{780},
  \bibinfo{pages}{5}.
\newblock \DOIprefix\doi{10.1088/0004-637X/780/1/5}.
%Type = Article
\bibitem[{{van de Hulst}(1950)}]{VdH50}
\bibinfo{author}{{van de Hulst}, H.C.}, \bibinfo{year}{1950}.
\newblock \bibinfo{title}{{The electron density of the solar corona}}.
\newblock \bibinfo{journal}{\bain} \bibinfo{volume}{11}, \bibinfo{pages}{135}.
%Type = Article
\bibitem[{{V{\'a}squez} et~al.(2009){V{\'a}squez}, {Frazin} and
  {Kamalabadi}}]{Vasquez09}
\bibinfo{author}{{V{\'a}squez}, A.M.}, \bibinfo{author}{{Frazin}, R.A.},
  \bibinfo{author}{{Kamalabadi}, F.}, \bibinfo{year}{2009}.
\newblock \bibinfo{title}{{3D Temperatures and Densities of the Solar Corona
  via Multi-Spacecraft EUV Tomography: Analysis of Prominence Cavities}}.
\newblock \bibinfo{journal}{\solphys} \bibinfo{volume}{256},
  \bibinfo{pages}{73--85}.
\newblock \DOIprefix\doi{10.1007/s11207-009-9321-1}.
%Type = Article
\bibitem[{{V{\'a}squez} et~al.(2010){V{\'a}squez}, {Frazin} and
  {Manchester}}]{Vasquez10}
\bibinfo{author}{{V{\'a}squez}, A.M.}, \bibinfo{author}{{Frazin}, R.A.},
  \bibinfo{author}{{Manchester}, IV, W.B.}, \bibinfo{year}{2010}.
\newblock \bibinfo{title}{{The Solar Minimum Corona from Differential Emission
  Measure Tomography}}.
\newblock \bibinfo{journal}{\apj} \bibinfo{volume}{715},
  \bibinfo{pages}{1352--1365}.
\newblock \DOIprefix\doi{10.1088/0004-637X/715/2/1352},
  \href{http://arxiv.org/abs/1012.5953}{{\tt arXiv:1012.5953}}.
%Type = Article
\bibitem[{{V{\'a}squez} et~al.(2011){V{\'a}squez}, {Huang}, {Manchester} and
  {Frazin}}]{Vasquez11}
\bibinfo{author}{{V{\'a}squez}, A.M.}, \bibinfo{author}{{Huang}, Z.},
  \bibinfo{author}{{Manchester}, W.B.}, \bibinfo{author}{{Frazin}, R.A.},
  \bibinfo{year}{2011}.
\newblock \bibinfo{title}{{The WHI Corona from Differential Emission Measure
  Tomography}}.
\newblock \bibinfo{journal}{\solphys} \bibinfo{volume}{274},
  \bibinfo{pages}{259--284}.
\newblock \DOIprefix\doi{10.1007/s11207-010-9706-1}.
%Type = Article
\bibitem[{{Zhang} et~al.(2005){Zhang}, {Ghodrati} and {Brooks}}]{Zhang05}
\bibinfo{author}{{Zhang}, Y.}, \bibinfo{author}{{Ghodrati}, A.},
  \bibinfo{author}{{Brooks}, D.}, \bibinfo{year}{2005}.
\newblock \bibinfo{title}{{An analytical comparison of three spatio-temporal
  regularization methods for dynamic linear inverse problems in a common
  statistical framework}}.
\newblock \bibinfo{journal}{Inverse Probl.} \bibinfo{volume}{21},
  \bibinfo{pages}{357--382}.
\newblock \DOIprefix\doi{10.1088/0266-5611/21/1/022}.

\end{thebibliography}

\end{document}